\documentclass[preprint2]{aastex}

\usepackage[usenames,dvipsnames]{xcolor}

\shorttitle{SFH of SMC fields}
\shortauthors{Cignoni et al.}

\begin{document}


%\title{Star formation history in the Small Magellanic Cloud: Four
%fields in the Wing, the main body and the outskirts\altaffilmark{1}}

\title{Mean age gradient and asymmetry in the star formation history
  of the Small Magellanic Cloud\altaffilmark{1}}
%\title{\textcolor{red}{More informative title?} Star formation history in the Small Magellanic Cloud:
%  the HST perspective\altaffilmark{1}}

\author{M. Cignoni\altaffilmark{2,3}, A. A. Cole\altaffilmark{4},
  M. Tosi\altaffilmark{3}, J. S. Gallagher\altaffilmark{5},
  E. Sabbi\altaffilmark{6}, J. Anderson\altaffilmark{6},
  E. K. Grebel\altaffilmark{7}, A. Nota\altaffilmark{6,8}}

\email{michele.cignoni@unibo.it}

%% affiliation.Based on observations with the
\altaffiltext{1}{Based on observations with the NASA/ESA Hubble Space
  Telescope, obtained at the Space Telescope Science Instutute, which
  is operated by AURA Inc., under NASAcontract NAS 5-26555. These
  observations are associated with program GO-10396.}
\altaffiltext{2}{Astronomy Department, University of Bologna, Bologna,
  40127, Italy} \altaffiltext{3}{INAF-Bologna Observatory, 40127,
  Italy}\altaffiltext{4}{School of Mathematics \& Physics, University
  of Tasmania, Private Bag 37, Hobart, Tasmania 7001, Australia }
\altaffiltext{5}{University of Wisconsin, Madison, WI,
  USA}\altaffiltext{6}{ STScI, Baltimore, MD, 21218, USA}
\altaffiltext{7}{Astronomisches Rechen-Institut, Zentrum f\"ur
  Astronomie der Universit\"at Heidelberg, M\"onchhofstr.\ 12--14,
  69120 Heidelberg, Germany }\altaffiltext{8}{European Space Agency,
  Research and Scientiﬁc Support Department, Baltimore, MD, USA}

\begin{abstract}

  We derive the star formation history in four regions of the Small
  Magellanic Cloud (SMC) using the deepest VI color-magnitude diagrams
  (CMDs) ever obtained for this galaxy. The images were obtained with
  the {\it Advanced Camera for Surveys} onboard the {\it Hubble Space
    Telescope} and are located at projected distances of 0.5--2
  degrees from the SMC center, probing the main body and the wing of
  the galaxy. We derived the star-formation histories (SFH) of the
  four fields using two independent procedures to fit synthetic CMDs
  to the data.
  
  We compare the SFHs derived here with our earlier results for the
  SMC bar to create a deep pencil-beam survey of the global history of
  the central SMC.  We find in all the six fields observed with HST a
  slow star formation pace from 13 to 5--7~Gyr ago, followed by a
  $\approx$2--3 times higher activity. This is remarkable because
  dynamical models do not predict a strong influence of either the
  Large Magellanic Cloud (LMC) or the Milky Way (MW) at that time.
  The level of the intermediate-age SFR enhancement systematically
  increases towards the center, resulting in a gradient in the mean
  age of the population, with the bar fields being systematically
  younger than the outer ones.  Star formation over the most recent
  500 Myr is strongly concentrated in the bar, the only exception
  being the area of the SMC wing. The strong current activity of the
  latter is likely driven by interaction with the LMC.
  
  At a given age, there is no significant difference in metallicity
  between the inner and outer fields, implying that metals are well
  mixed throughout the SMC.  The age-metallicity relations we infer
  from our best fitting models are monotonically increasing with time,
  with no evidence of dips.  This may argue against the major merger
  scenario proposed by \cite{tsujimoto2009}, although a minor merger
  cannot be ruled out.
\end{abstract}



\section{INTRODUCTION}

This paper is one of a series devoted to the derivation of the star
formation history (SFH) of the Small Magellanic Cloud (SMC) from the
comparison of deep, high resolution color-magnitude diagrams (CMDs) of
its resolved stars with synthetic CMDs based on stellar evolution
models.

It is part of a long-term project aimed at studying the evolution of
the SMC in space and time, using the Hubble Space Telescope (HST), the
Very Large Telescope (VLT), and the Guaranteed Time at the VLT Survey
Telescope (VST) to observe a large sample of field stars and clusters
across the SMC \citep[see,
e.g.,][]{nota2006,ripepi06,carlson07,sabbi07,glatt08a,glatt08b,tosi08,sabbi09,glatt11,cignoni12}. This
extensive data set will allow us to constrain the global SFH as well
as the existence of chemical and age gradients (if any) over the whole
lifetime and spatial extent of the SMC.
 
The SMC is an excellent benchmark to study the evolution of late-type
dwarf galaxies.  It is the closest dwarf irregular (dIrr), has a
current metallicity Z$\,\simeq 0.004$ (as derived from HII regions and
young stars) similar to that of the majority of dIrr's, and has a
mass, estimated between 1 and $5 \times 10^9\, M_{\odot}$, at the
upper limit of the range of this morphological class \citep[][and
references therein]{tolstoy09}.

  Thanks to the sensitivity and the large field of view of the VST, we
  will have CMDs a few magnitudes fainter than the oldest
  main-sequence (MS) turn-off (TO) for the entire galaxy and the
  Bridge connecting the SMC to the Large Magellanic Cloud (LMC).
  Meanwhile, we have already acquired deeper, higher-resolution
  photometry of 4 SMC young clusters \citep{nota2006}, 7
  intermediate-age and old clusters \citep{glatt08a,glatt08b} and 6
  fields \citep{sabbi09} with the Advanced Camera for Surveys (ACS) on
  board of HST.

  We have chosen the six HST fields to sample regions characterized by
  different star formation activity, stellar and gas densities. They
  are located in the SMC bar, in the outskirts, and in the wing,
  i.e. the large cloud of faint stars protruding toward the LMC, that
  is considered to be part of a tidal tail torn off the main body of
  the SMC by the interactions between the two Clouds \citep{wes64}.

\begin{figure*}
\centering 
\includegraphics[width=13cm,angle=-90]{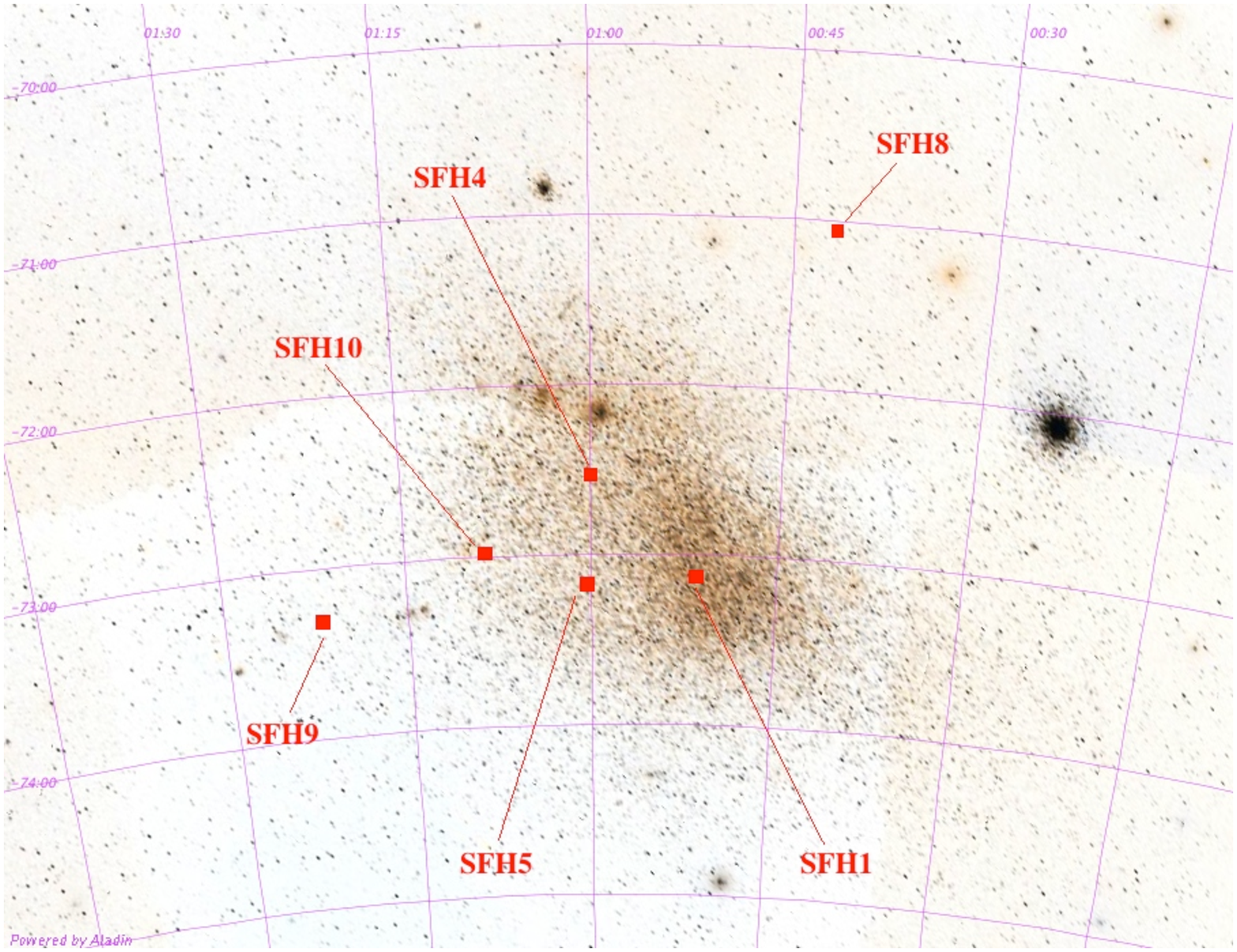}
\caption{Spatial distribution of our 6 SMC fields observed with
  HST/ACS, superimposed on the DSS image of the SMC. North is up, and
  east is left.}
\label{fields} 
\end{figure*}

  We derive the SFHs of the observed fields using the synthetic CMD
  technique \citep[see e.g.][and references
  therein]{tolstoy09,cignoni10}. To estimate the intrinsic theoretical
  uncertainties, the SFH is derived using two completely independent
  procedures for the application of the synthetic CMD method: the
  Bologna code (see e.g. \citealt{cignoni10}) and Andrew Cole's
  annealing procedure \citep{cole07}.  We have summarized the two
  methods, their commonalities and differences, in the paper by
  \cite{cignoni12}, where we applied them to the two most central HST
  fields of our sample of six. Those two fields are located in the SMC
  bar. Here we present the results for the remaining four HST fields:
  one (SFH10) between the bar and the wing, one (SFH9) in the wing,
  one (SFH5) in the ``central system'' \citep{wes97} but away from the
  bar, and one (SFH8) in the galaxy outskirts, at the same distance
  from the center as our wing field SFH9 but in the direction opposite
  to the LMC (see Fig. \ref{fields}).
%\clearpage

SFHs of other SMC fields have been derived and presented by other
groups, based on HST images of small individual regions or on
ground-based photometry \citep[see Fig. 1
in][]{cignoni12}. \citet{gard92} analyzed UKST plate data totalling
130 square degrees around the SMC; this remains the largest areal
coverage CMD analysis published to date, but does not reach below the
horizontal branch/red clump.  By comparison, CCD studies have covered
much smaller areas. \citet{harris04} derived the SFH of the SMC over
$4\degr \times 4.5\degr$ to a depth of $V\la 21$ using the Magellanic
Cloud Photometric Survey (MCPS) UBVI catalog by
\citet{zaritsky2002}. Recently, \citet{nidever2011} published the
first results from a $\approx$15 deg$^2$ survey covering selected
fields at angular distances of 2--11$^{\circ}$ from the SMC center
defined by \cite{mateo98}, deep enough to reach the old MSTO in the
uncrowded outer regions.  \citet{noel07} and \cite{noel09} presented a
deep ground-based study of 12 fields of the SMC, avoiding the densest
regions, because of their high crowding conditions.

HST studies have usually concentrated on regions of recent star
formation and/or high crowding. \citet{dolphin01} analyzed the stellar
content at the outskirts of the SMC, in a region close to the globular
cluster NGC~121, using both HST Wide Field Planetary Camera 2 (WFPC2)
and ground based data.  \citet{mccumber05} studied the stellar content
of a small portion of the SMC wing also with the WFPC2.  A summary of
WFPC2 studies and reanalysis of the CMD data has been undertaken by
\citet{weisz13}.  With ACS, \cite{chiosi07} have derived the SFH in
the vicinity of a few SMC clusters. Here, we perform a quantitative
analysis of the fields located at a range of radial distances from the
SMC center, described in Sabbi et al.\ (2009).

The layout of this paper is as follows: the HST data and resulting
CMDs are briefly described in Section 2, while the SFH derivation with
the two different procedures is presented in Section 3. Similarities
and differences between the resulting SFHs are also discussed
there. In Section 4 we compare our results with published literature
on other SMC regions. Our conclusions close the paper.

\section{HST Photometry and CMDs}

Our six SMC fields were imaged in the F555W and F814W filters with the
ACS Wide Field Channel (WFC) from November 2005 and January 2006
(GO-10396; P.I. Gallagher). The data reduction was performed with the
program img2xym\_WFC.09x10 \citep{anderson06}, and the resulting
magnitudes were calibrated in the Vegamag photometric system using
\citet{sirianni05} prescriptions. For sake of conciseness, from now on
we will refer to the $m_{F555W}$ and $m_{F814W}$ magnitudes calibrated
in the Vegamag system as V and I, respectively.

Extensive artificial star experiments were performed following the
approach described in \citet{anderson08} to test the level of
completeness and the photometric errors of the data. The artificial
stars were searched for with exactly the same procedure adopted for
the real stars. We considered an artificial star recovered if its
input and output positions agreed to within 0.5 pixels and the fluxes
agree to within 0.75 mag. As done for the photometric analysis we also
required that each star is found in at least three exposures with a
positional error $<0.1$ pixels per filter.  Details on both the
photometry and the artificial star tests can be found in
\cite{sabbi09}.

\begin{figure*}
\centering
\includegraphics[width=15cm]{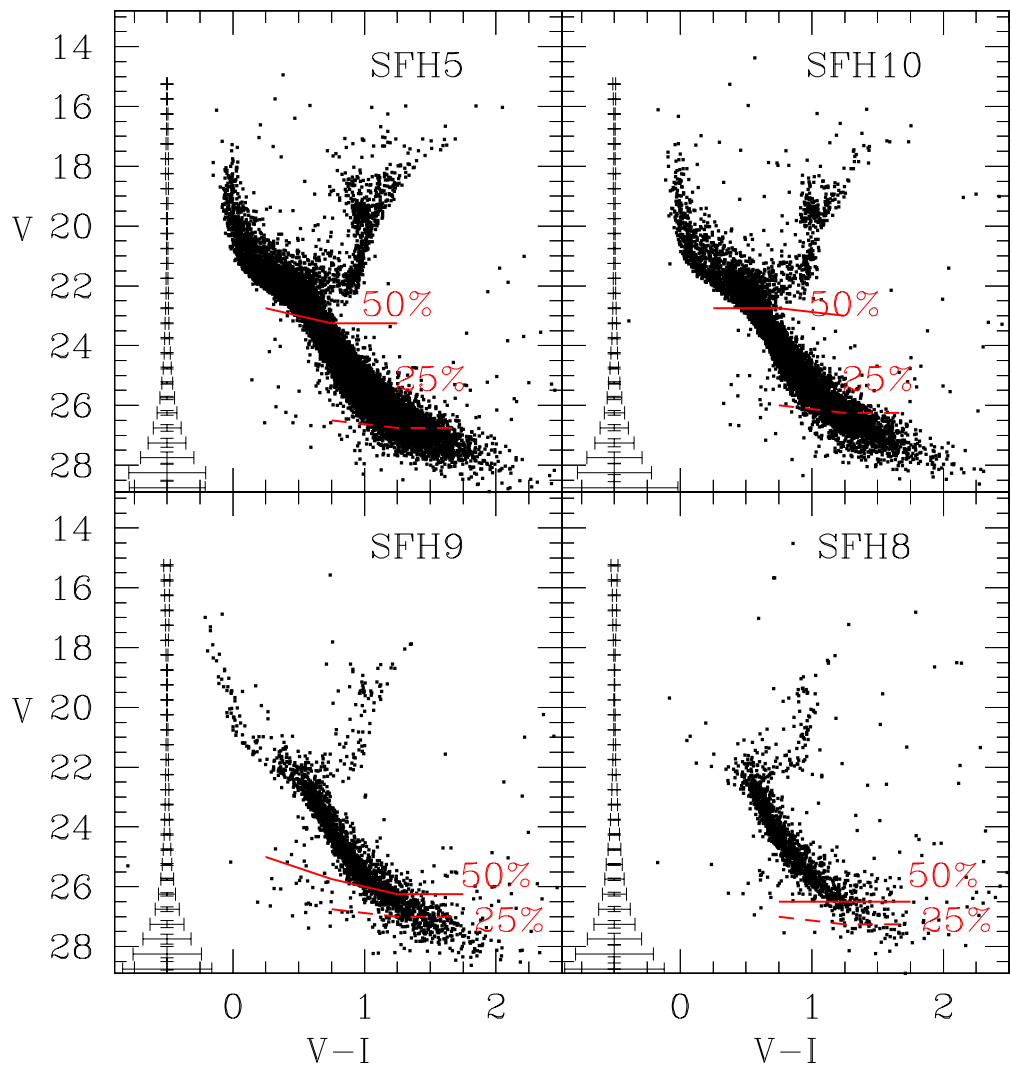}
\caption{CMDs of the SFH5, SFH8, SFH9 and SFH10  fields
  observed with the ACS/WFC. The solid and the dashed red lines
  indicate the 50\% and 25\% levels of completeness, respectively.
  Formal errors on the estimated photometry are shown on the left side
  of each CMD (see text for details).}
\label{cmdobs} 
\end{figure*}

The final CMDs are shown in Fig.\ref{cmdobs}, where the photometric
errors are also plotted\footnote{To be conservative, the plotted error
  bars correspond at each magnitude level to the larger value of the
  error resulting from the photometric package and from the artificial
  star tests (which tend to be slightly larger than those based on
  point-spread-function photometry alone). }. These CMDs show the
superb tightness typical of HST photometry and reach almost four
magnitudes fainter than the oldest MSTO, thus allowing us to study the
evolution of the regions over the whole Hubble time.

All the four CMDs display well populated sequences of all the
evolutionary phases of old and intermediate-age stars: MS, sub-giant
branch (SGB), red giant branch (RGB) and red clump (RC).  In addition
to these components, all the CMDs except that of SFH8 present a bright
blue plume typical of young, high- and intermediate-mass stars. None
of the CMDs, with the possible exception of SFH9, shows any
significant population of pre-main sequence (PMS) stars. This suggests
a star formation activity in the last 50 Myr much lower than that of
the currently most active regions of the SMC, such as NGC~346 and
NGC~602, where we measured many PMSs with the same observing setup
(see, e.g., \citealt{nota2006, carlson07, cignoni09,
  cignoni10b,cignoni11}).  None of the CMDs show evidence of a
horizontal branch (HB), suggesting that in all the fields the star
formation activity was quite low at epochs earlier than around 10 Gyr
ago.

The CMDs differ from each other for some key features. First of all,
the number of stars present in the CMD strongly depends on the
apparent galactocentric distance of the region. The final photometric
catalogs contain about 29200 objects in SFH1, 17300 in SFH4, 19770
objects in SFH5, 1560 in SFH8, 2660 in SFH9, and 9180 in SFH10
\citep{sabbi09}. In practice, the most populated region is the most
central one (SFH1). SFH4 and SFH5, with a similar projected distance
from the SMC center, contain a similar number of stars, and the more
external fields host much fewer objects. 

The bar field SFH1 is the closest ($\sim 3\arcmin$) to the SMC optical
center defined by \cite{gonidakis09}\footnote{This center is found
  using K- and M-stars, hence provides a good estimate of the stellar
  mass distribution.} and contains $\sim8.5\, {\rm stars\,
  pc}^{-2}$. SFH4 is also in the bar, at $\sim 53\arcmin$ from the SMC
optical center and has a stellar density of $\sim 5\, {\rm stars\,
  pc}^{-2}$.  SFH5 is located at $\sim 41\arcmin$ from the center, but
at right angles to the bar major axis; its density of $\sim 5.7\, {\rm
  stars\, pc}^{-2}$ is second only to the density in SFH1, emphasizing
that while the bar is prominent in the younger populations, the older
stars are more symmetrically distributed (see e.g. \citealt{gard92}
and Figures 2 and 3 in \citealt{zaritsky2000}). SFH8 is $\sim 2\degr
10\arcmin$ North from the SMC optical center, and has the lowest
stellar density of all our HST fields, $\sim 0.5\, {\rm stars\,
  pc}^{-2}$, almost a factor of 20 lower than in SFH1. SFH9 is located
in Shapley's wing, and is the region most distant in projection from
the optical center of the SMC ($2\degr 14\arcmin$) in our sample. Its
stellar density is as low as $\simeq 0.8\, {\rm stars\, pc}^{-2}$.
Midway between the bar and wing, SFH10 lies $1\degr 17\arcmin$ from
the center. Its stellar density is $\simeq 2.7\, {\rm stars\,
  pc}^{-2}$, 3.5 times denser than SFH9 and 3.1 times less dense than
SFH1, in good agreement with the slope of the surface brightness fit
by \citet{bothun88}.

In terms of populations there is an obvious age difference between
SFH8 and the others: it shows no blue plume of MS stars brighter than
V $\approx$21, even though it is well within the boundary of the area
described by \citet{nidever2011} as the ``inner component'' and by
\citet{wes97} as the ``central system'' (radius 3--3.5$^{\circ}$).  In
contrast, SFH9, although at an apparent slightly larger distance from
the center, has a very narrow and well-defined blue plume, bluer at
its bright end than any of the others. This is the locus of very young
stars, and implies that the most external parts of the wing are still
forming stars, thus emphasizing the asymmetry of the activity in the
SMC. The narrowness of the upper MS in SFH9 compared to the broad blue
plumes in the more central fields emphasizes the likelihood that the
wing represents a single star formation event and not a spatial
redistribution of the populations in the central regions.

\section{SFH and Age-Metallicity Relation of the four fields}

The SFHs of SFH5, SFH8, SFH9 and SFH10 have been derived with the
synthetic CMD method following two independent procedures: Cole's
\citep[e.g.][]{cole07} and Bologna's, the latter being a combination
of the procedure developed by Cignoni \citep[e.g.][]{cignoni06a, cignoni06b} with
those defined and improved over the years at the Bologna Observatory
\citep[see][]{tosi91,greggio98,angeretti05}.  Commonalities and
differences of Cole's and Bologna's methods were summarized by
\cite{cignoni12}.

For these four fields we followed exactly the same procedures and
assumptions as described in the latter paper for SFH1 and SFH4. In all
cases the synthetic CMDs have been built to reproduce the number of
stars measured in the observational diagrams, using the results of the
artificial star tests, described in the previous section, to assign
photometric errors and incompleteness to the synthetic stars. The
synthetic stars are simulated from the stellar evolution models
computed by the Padova group \citep{bertelli08,bertelli09} and
converted directly to the HST Vegamag photometric system, to minimize
the uncertainties related to calibration issues.

Initial Mass Function (IMF), binary fraction, reddening, differential
reddening and distance modulus in principle are allowed to vary
freely, but in practice always turn out to be viable only within
restricted ranges of value. No age-metallicity relation (AMR) is
assumed a priori.

The best solution is searched in a statistical manner ($\chi^2$
minimization over appropriate CMD grids with a downhill simplex
algorithm in the Bologna case, and a simulated annealing approach for
maximum likelihood based on a \citealt{cash79} statistics in Cole's). The
quantitative solutions are not truly unique although the optimization
methods are both highly tuned to produce nearly-formally unique
results. By showing the results from both methods, we allow for a
robust derivation of the range of parameter values that is more
realistic than the errorbars resulting from any single method.

Distances and reddenings are initially constrained to the values given
in \citet{sabbi09}, but are allowed to vary if the resulting synthetic
CMDs do not optimally match the data.  All the parameters of the SFH
solutions for each field are summarized in Table~\ref{tab-models}.
\begin{table*}[]
\footnotesize
\begin{center}
\caption{Summary of SFH Solution Parameters$^a$}\label{tab-models}
\begin{tabular}{lccccccc}
\hline
Field & Method & (m$-$M)$_0$ & E(B$-$V) & IMF & CMD binning & Metallicities \\
         &        &    (mag)      &  (mag)   &     & (color$\times$mag) & (Z$\times$10$^3$) \\
\hline
SFH5   & Bologna & 18.85& 0.08$^b$& Kroupa (2001)   &  variable &0.4,1, 2, 4 \\
  Central & Cole    & 18.95 & 0.07$^b$ & Chabrier (2003) &  0.04$\times$0.08 & 0.15, 0.4, 0.6, 1.0, 1.5, 2.4, 4.0\\
\hline
SFH10   & Bologna & 18.83& 0.095$^b$& Kroupa (2001)   &  variable & 0.4, 1, 2, 4\\
 Intermediate  & Cole    & 18.89 & 0.06$^b$ & Chabrier (2003) &0.04$\times$0.08 & 0.15, 0.4, 0.6, 1.0, 1.5, 2.4, 4.0\\
\hline
SFH9   & Bologna & 18.81& 0.10& Kroupa (2001)   &  variable & 0.4, 1, 2, 4\\
 Wing & Cole    & 18.90 & 0.10 & Chabrier (2003) &0.04$\times$0.08 & 0.15, 0.4, 0.6, 1.0, 1.5, 2.4, 4.0\\
\hline
SFH8   & Bologna & 18.85& 0.087& Kroupa (2001)   &  variable & 0.4, 1, 2, 4\\
 Outer  & Cole    & 18.96 & 0.06 & Chabrier (2003) & 0.04$\times$0.08 & 0.15, 0.4, 0.6, 1.0, 1.5, 2.4, 4.0\\
\hline
\end{tabular}
\medskip\\
\end{center}
$^a$All models based on the Padova isochrone set; see text for
details.
$^b$Differential reddening assumed for stars younger than 500 Myr; see text
for details.
\end{table*}

The best-fit distances always correspond to distance moduli shorter
than recently determined from eclipsing binaries ($(m-M)_0=19.11$,
\citealt{north2010}), but still compatible with the average distance
of RR-Lyrae ($(m-M)_0=18.90$, \citealt{kapakos2011}) and of star
clusters (around 18.87 for \citealt{glatt08b} and between 18.71 and
18.82 for \citealt{crowl2001}). These differences may be, at least
partially, due to the line of sight depth variations found by
\citealt{subra2009} (up to 4.9 kpc) and \citealt{glatt08b} (between 10
and 17 kpc).  Indeed, the old SMC population (as traced by RR Lyrae
stars) has a mean depth of $4.2 \pm 0.4$ kpc (with maximum values up
to 5.6 kpc).  The young populations (as traced by Cepheids) not only
show a radically different distribution, but also higher depths: in
total, a mean depth of $7.5 \pm 0.3$ kpc.  In particular, several of
the regions in which our HST fields are located belong to areas with
very large line-of-sight depth values (see Figure 6 in
\citealt{haschke12}). In addition, the orientation of the spatial
distribution of the young SMC population is such that the northeastern
part (where several of our HST fields are located) is closer to us
than the rest (see \citealt{haschke12}).

Note that in each field the SFH solutions from the two
different methods find systematically different distances, although
the differences ($\approx$0.1 mag $=$ $\approx$3~kpc) are within both
the likely errors and the physical depth of the cloud.

The best fitting reddenings are in good agreement with the foreground
value found by \cite{schlegel98}.  To better reproduce the color width
of the upper MS, the solutions for SFH5 and SFH10 required a small
amount of differential reddening: the Bologna solutions added an
additional E(B$-$V) $=$0.02 for stars younger than 500 Myr while Cole
added 0.02 to SFH5 and 0.03 to SFH10, for stars younger than 250~Myr.
Indeed, reddening has been demonstrated to be highly variable and
differential in the SMC \citep[see e.g.][]{zaritsky2002, haschke11}.

In the next sections Cole's and Bologna best-fit solutions are
presented as SFH\#-C and SFH\#-B respectively (e.g. SFH5-C indicates
Cole's SFH for SFH5) and compared with each other. Since SFH5 is the
only field where Bologna and Cole solutions differ more than formal
errors, for this region we will discuss the CMD residuals in some
detail. For the other fields, where the resulting SFHs are well
consistent within the uncertainties, we will show only the synthetic
CMD corresponding to Bologna's solutions.

%\clearpage
\subsection{SFH5  \label{sec-sfh5}}

\begin{figure*}
\centering
\includegraphics[width=12cm]{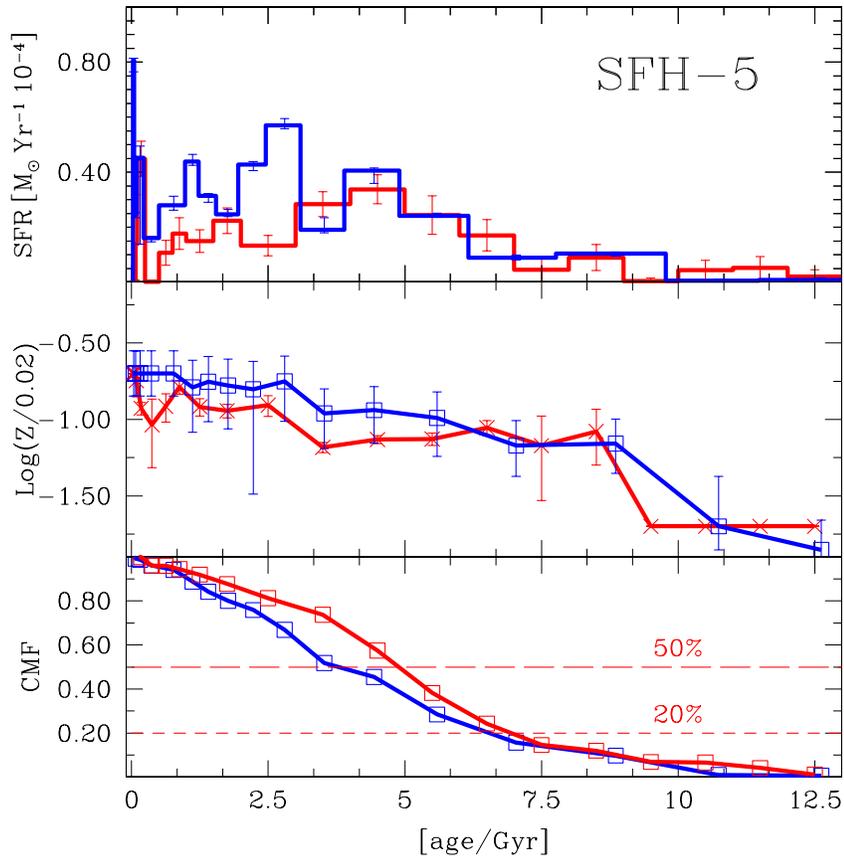}
\caption{Recovered SFHs (top panel), AMRs (middle panel) and CMF
  (bottom panel) for SFH5 using the Bologna (SF5-B, red histogram) and
  Cole procedure (SFH5-C, blue histogram).}
\label{sfr5} 
\end{figure*}

SFH5 is the densest region of the four examined here and the second
densest of all our six HST fields. In its CMD we find the signatures
of both very old stars (the fainter SGB stars are at least 12 Gyr old)
and fairly young ones (the bright blue plume). When simulated with the
synthetic CMD method, this region appears to have experienced a mildly
{\it gasping} \citep[originally defined by][]{marconi95} regime of
star formation over most of its life, with peaks and dips of similar
duration and rates within a factor of two of the average value. The
top panel of Figure \ref{sfr5} shows SFH5's SFH as recovered using
Bologna (SFH5-B, red line) and Cole's procedures (SFH5-C, blue line).

\begin{figure*}
\centering
\includegraphics[width=14cm]{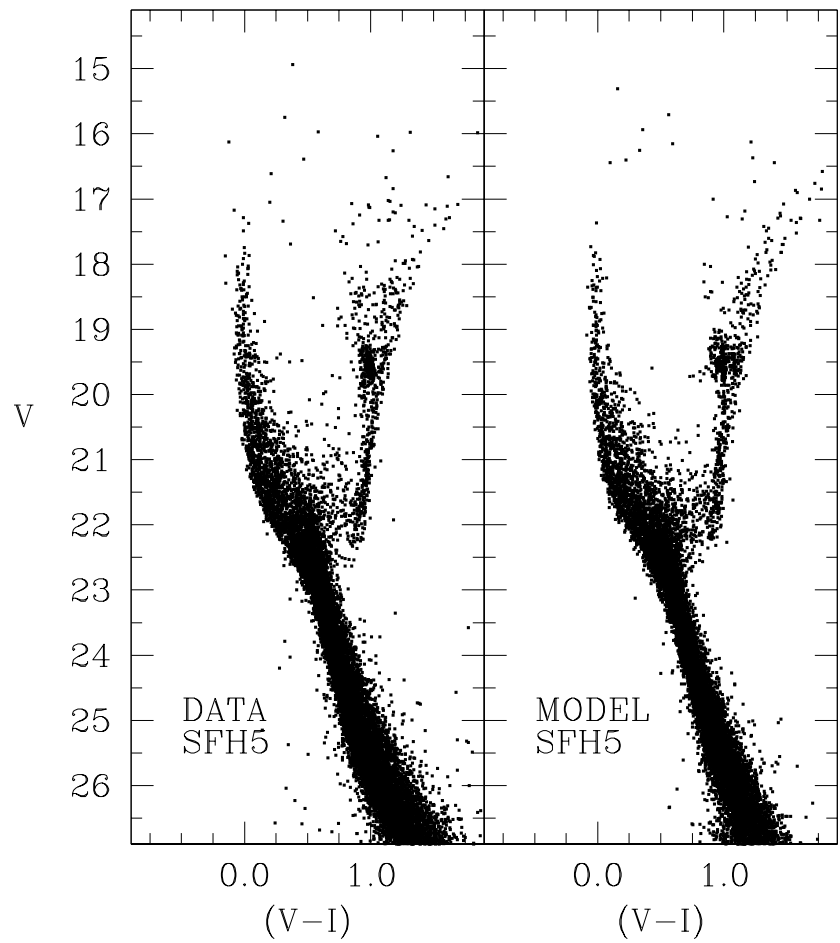}
\caption{Comparison between the observational (left panel) and
  the Bologna synthetic CMD (right panel) for SFH5. }
\label{cmd_5} 
\end{figure*}

Figure \ref{cmd_5} shows the synthetic CMD (right panel) drawn from
the Bologna SFH compared to the data (left panel). From a
morphological point of view, the MS and the SGB are well reproduced,
while the RC and Blue Loop (BL) region are, respectively, slightly
broader and less populated than the observational ones.

As shown in Fig. \ref{sfr5}, the two solutions share the common
characteristics of a significant discontinuity between the activity in
the earliest 5-6 Gyr and the subsequent epochs. Stars were already
being formed in the earliest phases, but at a very modest rate,
significantly lower than at later times.

Overall, both solutions predict that: 1) only 20\% of the stellar mass
was in place before 7.5~Gyr ago (see bottom panel of Fig. \ref{sfr5},
where the cumulative mass fraction, CMF is plotted); 2) The average SFH
has not dropped significantly since its major event at 5~Gyr ago.

However there are apparent differences between the two results. First,
the degree of ``burstiness'' is higher in Cole's solution than in
Bologna's. SFH5-C shows four SFR peaks a factor of two above the
average: 4--5~Gyr ago (secondary peak), 2--3~Gyr ago (primary peak),
1~Gyr ago and in the last 250 Myr, while SFH5-B shows a relatively
smooth recent history, with only two peaks slightly above the average,
at about 4.5 Gyr ago (primary peak) and 1.5--1.8~Gyr ago with gasps
$\approx$0.4 and 1.2~Gyr. These differences can be partially explained
by the higher age resolution adopted in Cole's approach. Second, the
CMFs are slightly different in the range 2.5-5 Gyr ago, with SFH5-B
reaching the 50\% level about 1 Gyr before SFH5-C. This effect is
likely due to a combination of several model inputs, including the
adopted metallicity grid\footnote{Although the two codes adopt the
  same Padova library, Bologna's code uses only the stellar tracks
  provided by \citep{bertelli08,bertelli09}, Cole's code uses also
  tracks with interpolated metallicities.}, the CMD binning and
interpretation (Bologna's approach attempts to fit the whole CMD,
whilst Cole's approach restricts the analysis to MS and SGB stars),
the general ability of each approach to escape from local minima,
etc. Since these effects are strongly interlaced, we consider the
differences in the solution as a measure of the systematic
uncertainty.

\begin{figure*}
\centering
\includegraphics[width=5.2cm]{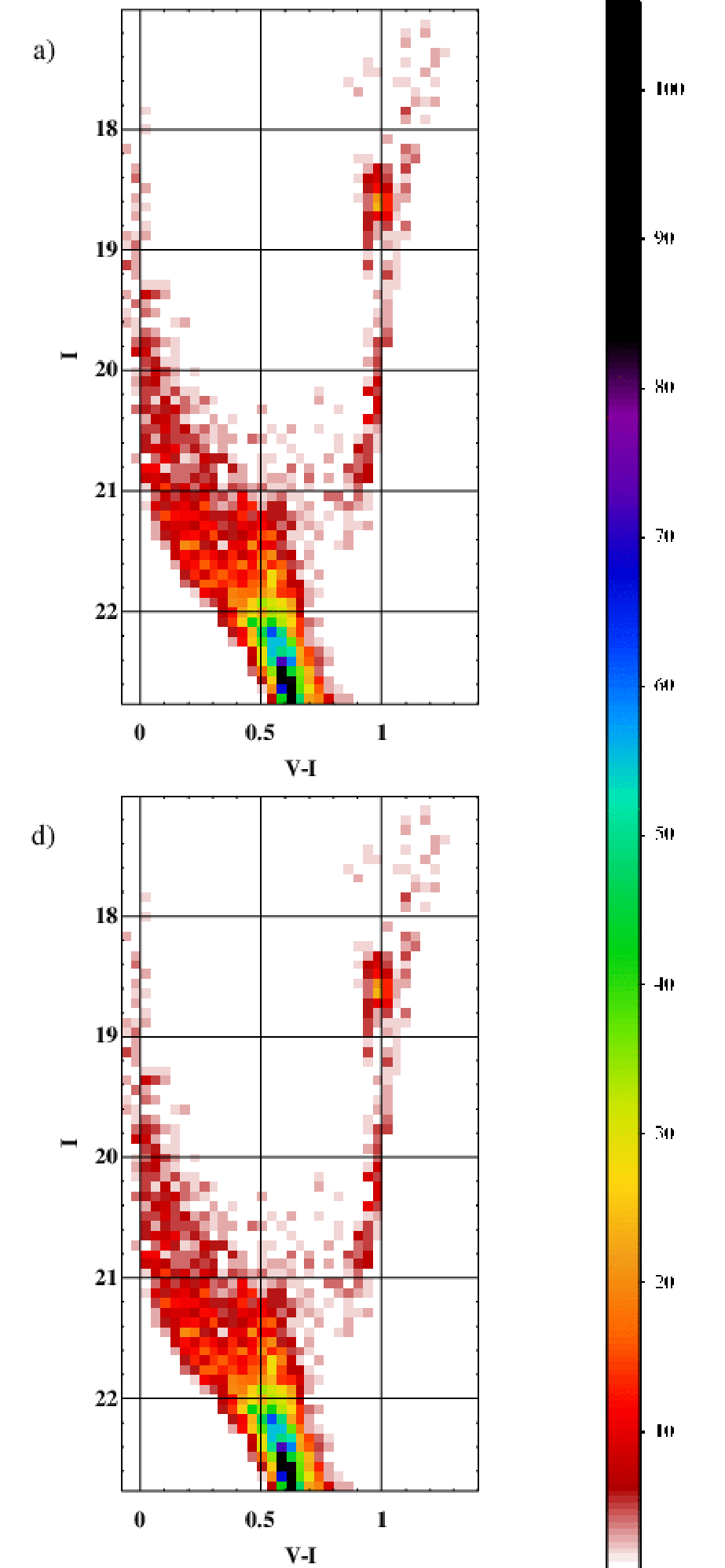}
\includegraphics[width=5.2cm]{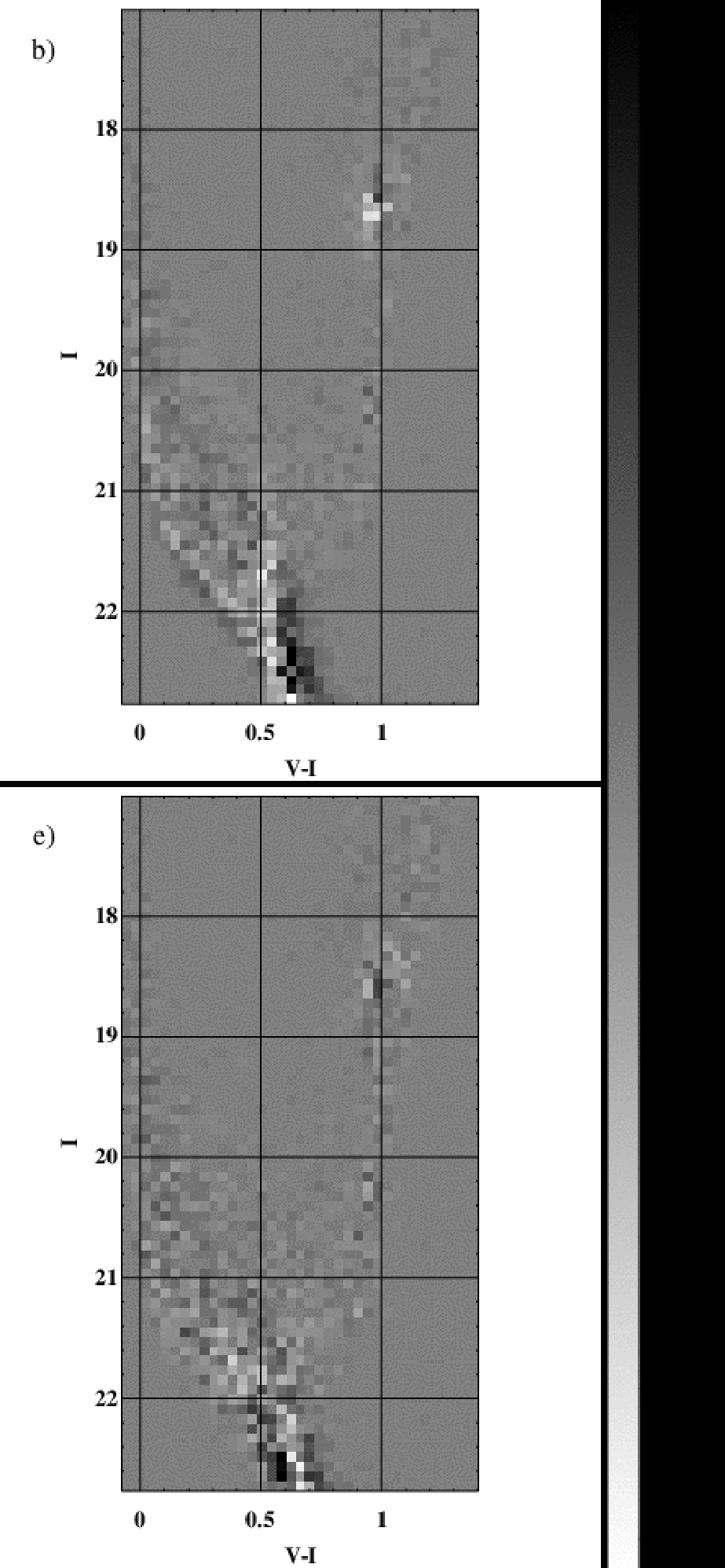}
\includegraphics[width=5.2cm]{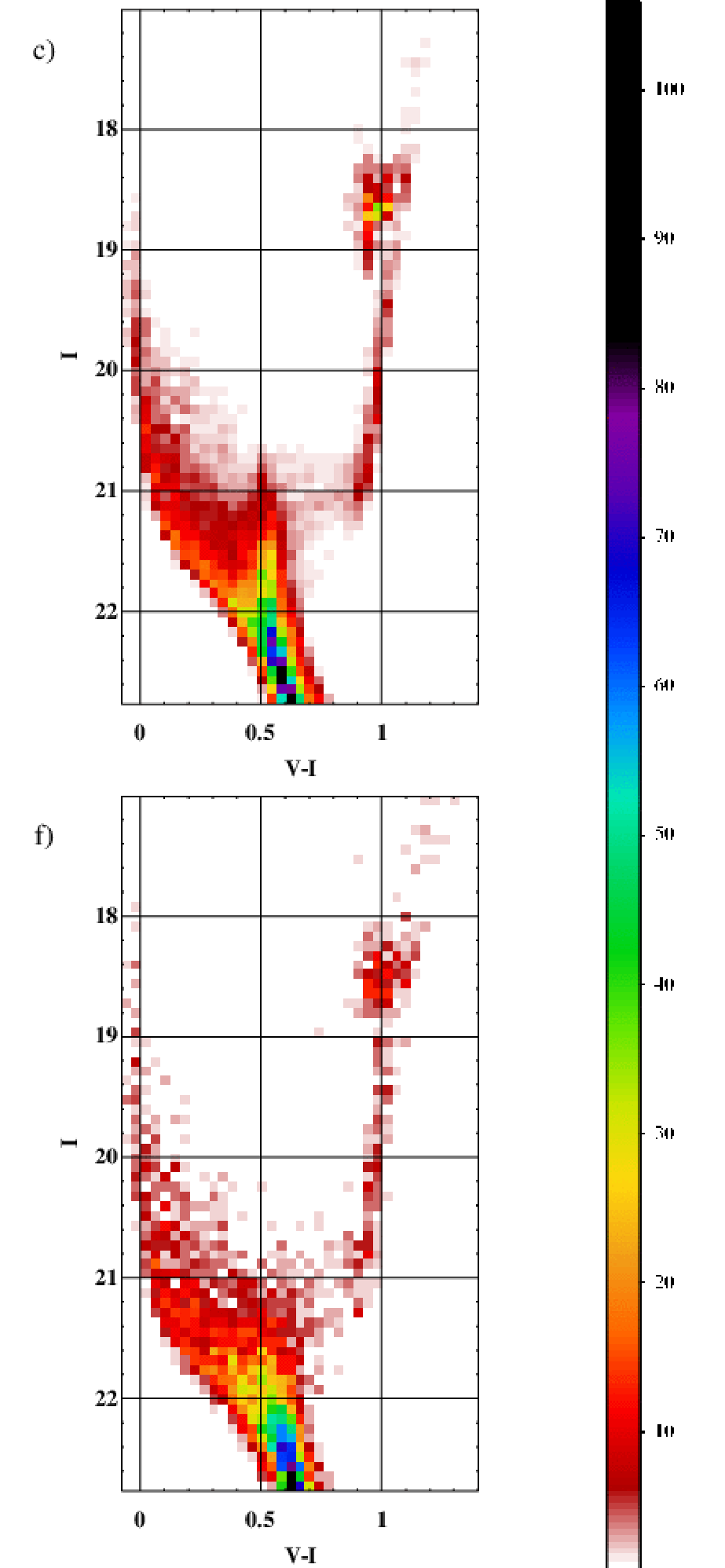}
\caption{Comparison between observational and synthetic Hess diagrams
  for SFH5. Panels [a] and [d] show the data; panels [b] and [e] show
  the Cole and Bologna residuals (subtraction of the observed Hess
  diagram from the calculated one); panels [c] and [f] show the Cole
  and Bologna best models, respectively.}
\label{hess_5} 
\end{figure*}

To compare the performances of the two solutions, we plot in
Fig. \ref{hess_5}, from left to right, the Hess diagram for the data
(panels [a] and [d]), residuals (panel [b] for Cole, panel [e] for
Bologna), best model CMDs (panel [c] for Cole and panel
[f] for Bologna).
From these diagrams it can be seen that both models show equally
scattered residuals in the regions of the MS (above I=22), SGB and
RGB, with no evident systematic departure (except for a vertical mismatch
of Cole's model at $V-I\approx 0.6$). On the other hand, both models
have shortcomings in reproducing details of the RC region: Cole's
model predicts too many RC stars, but qualitatively reproduces the RC
morphology; Bologna's model fits well the number of RC stars, but
predicts a slightly different shape of the RC.

\begin{figure*}
\centering
\includegraphics[width=8cm]{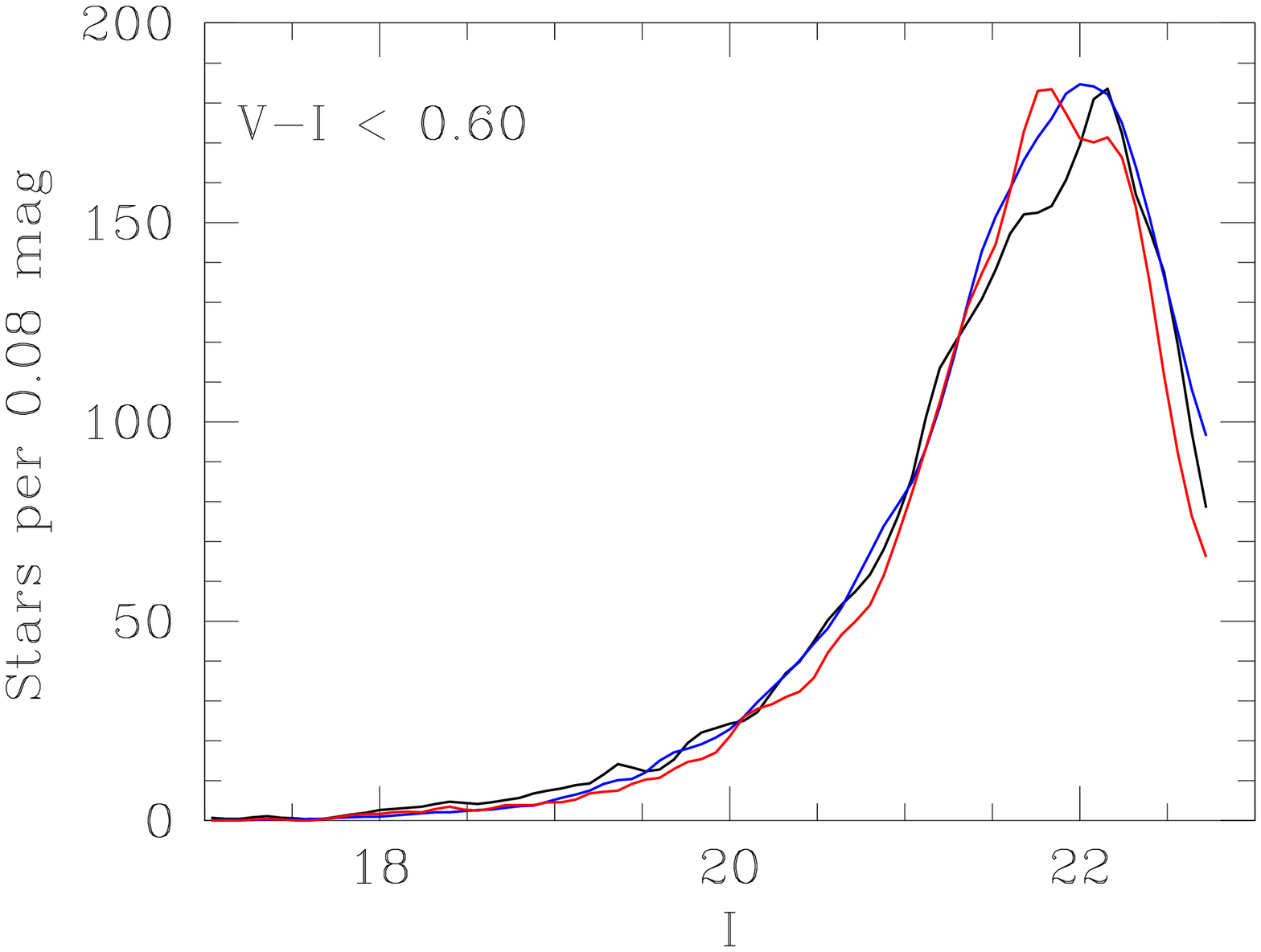}
\includegraphics[width=8cm]{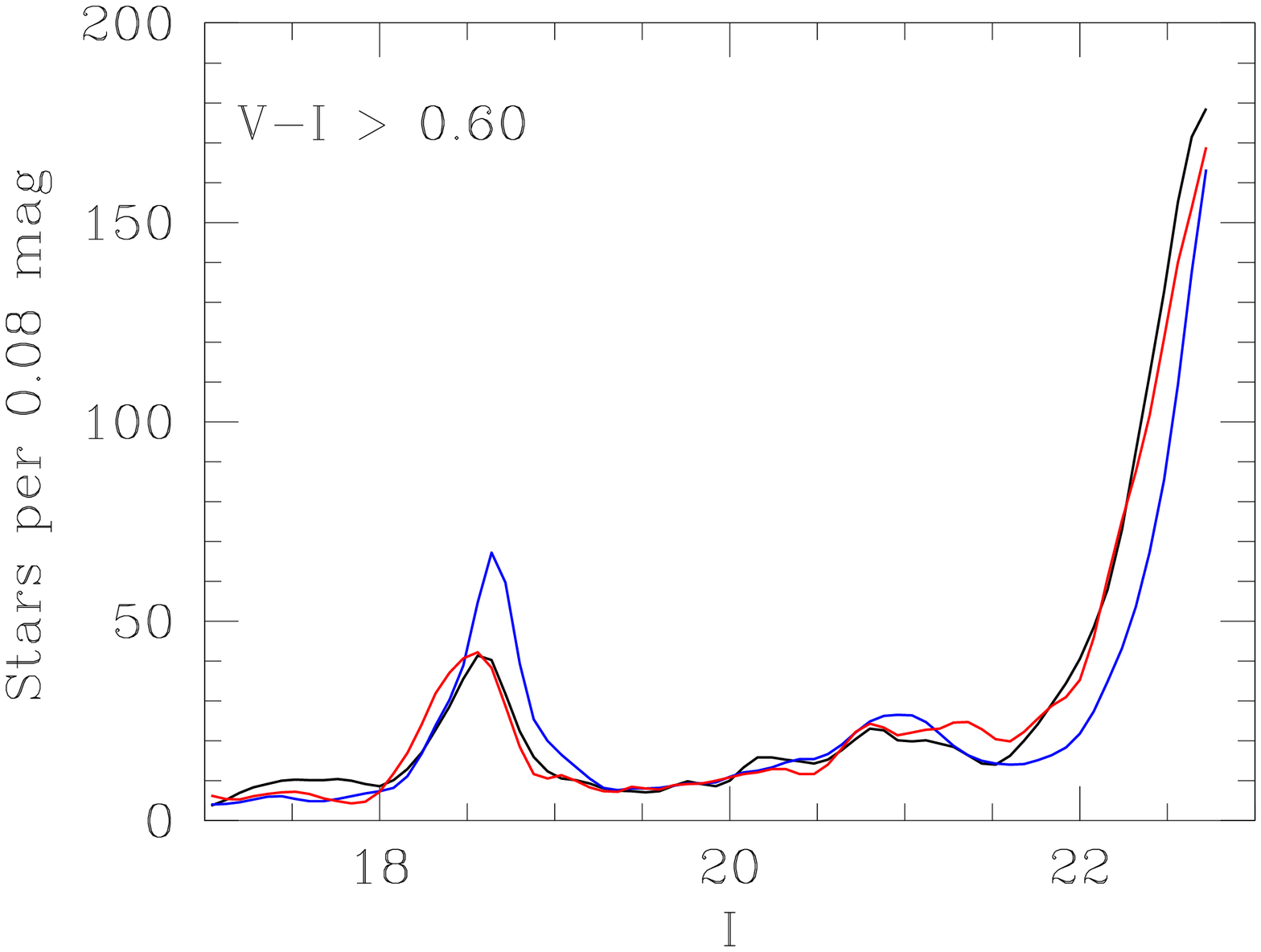}
\caption{Comparison of the predicted and observational luminosity
  functions (I mag) for blue (left panel) and red (right panel) stars
  in SFH5. The blue lines correspond to Cole's best solutions, the red
  lines to Bologna's, and the black lines to the data.}
\label{LF} 
\end{figure*}

More subtle differences appear when one compares the luminosity
functions (LFs, Fig. \ref{LF}) from our models with the observational
ones. In the left panel (stars with $V-I<0.6$) the Bologna model
slightly underpredicts the number of MS stars in the magnitude range
$20<I<21$, and both models overpredict the observed counts in the MS
range $21.5<I<22$. In the right panel (stars with $V-I>0.6$), Cole's
model clearly overpopulates the RC (the peak around I=18.6). The
latter mismatch is a likely consequence of Cole's approach which uses
only MS and SGB stars to recover the SFH, leaving RC and RGB regions
unconstrained.

Concerning the recovered chemical history, Bologna's metallicity is
slightly higher than Cole's prior to 7.5 Gyr ago, while the opposite
is true at later times. However, the two results are consistent with
each other within the uncertainties, also considering the coarser bin
size (0.3 dex) of the Bologna metallicity network.

The differences between the Bologna and Cole solutions are the result
of the assumptions of the codes and of the way in which the best-fit
models are determined.  The normalization over the past $\approx$5~Gyr
(Cole's SFR is slightly higher than Bologna's) can, at least in part,
be attributed to the different assumed IMFs, which diverge at
1~M$_{\odot}$, roughly the MSTO mass for a 5~Gyr age at SMC
metallicities. The differences in bursty vs.\ smooth SFH are likely
the result of degeneracies between age, metallicity, distance, and
reddening, and the way in which those degeneracies are broken given
the adopted isochrones. Age-metallicity effects are suspected here
owing to the divergence between the inferred AMRs from 1--3~Gyr ago,
where the SFH differences are most pronounced.

Of the four fields analyzed in this paper, only SFH5 shows differences
larger than the formal uncertainties on the solutions; it is no
coincidence that this is the most crowded field, with the highest SFR,
where differential reddening is also present. Similar effects were
seen at a similar level in the bar fields analyzed in Paper~I.

The overall SFH in SFH5 is qualitatively similar to those of our bar
fields SFH4 and SFH1 (see \citealt{cignoni12}); all three fields
formed the majority of their stars in the last 5--6 Gyr. The very low
rate of early star formation ($>$ 10 Gyr ago) is consistent with the
lack of a net HB in their CMDs. Additionally, the average SFR has been
almost constant (SFH1) or slightly declining (SFH5 and SFH4) over the
past few Gyr, although peaks and dips at various times and amplitudes
are observed, as typical of gasping regimes.

\subsection{SFH8  \label{sec-sfh8}}

\begin{figure*}
\centering
\includegraphics[width=12cm]{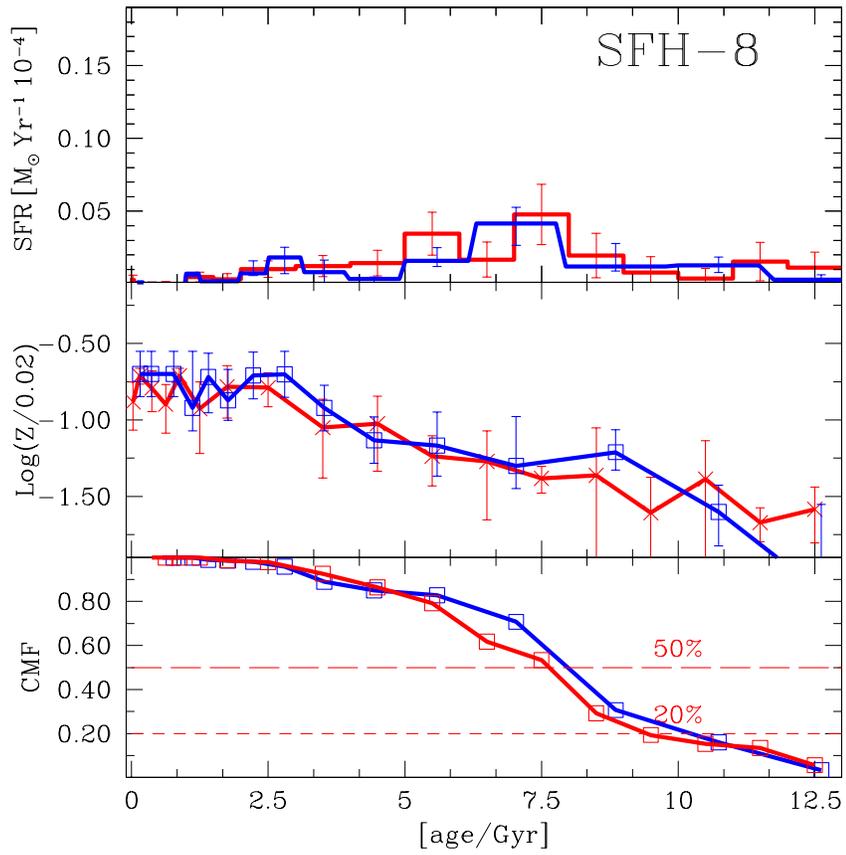}
\caption{SFH, AMR and CMF for SFH8. Colours and lines are as in Figure
  \ref{sfr5}.}
\label{sfr8} 
\end{figure*}

\begin{figure*}
\centering
\includegraphics[width=12cm]{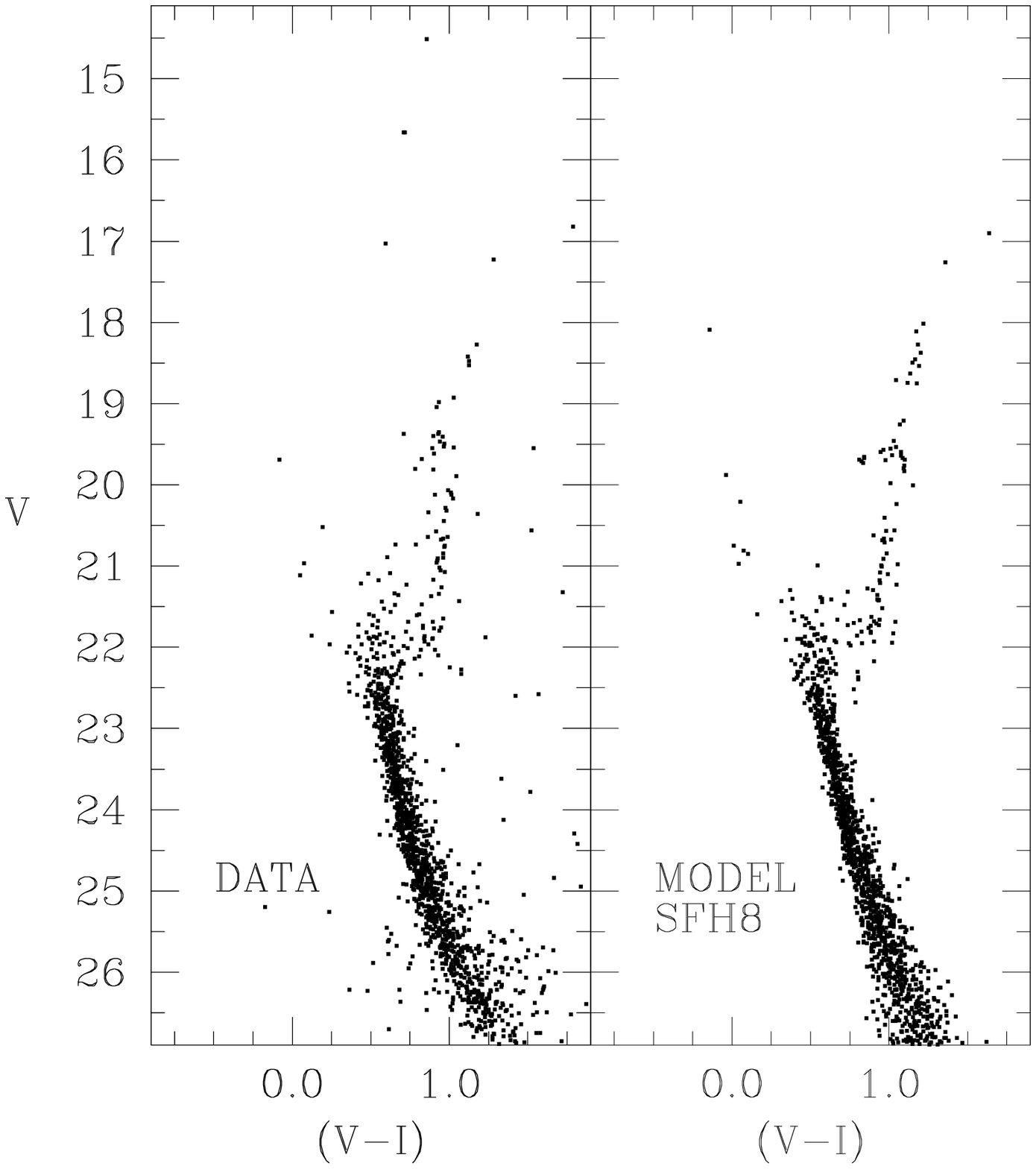}
\caption{Comparison between the observational (left panel) and
  the Bologna synthetic (right panel) CMD for SFH8.}
\label{cmd_8} 
\end{figure*}

SFH8 is our most distant field in the SMC outer regions. Its CMD is
characterized by the lack of a clear blue plume, as expected for a
region sufficiently away from the star forming body of the
galaxy. Although it lies well within the density profile break taken
to mark the transition to a ``halo'' \citep{nidever2011}, it shows no
evidence for recent star forming activity. Both the solutions SFH8-B
and SFH8-C (see top panel of Figure \ref{sfr8}) find that the field
has been quiescent since $\sim 1$ Gyr. Both methods also agree on a
rather moderate SF activity earlier than 7 Gyr ago, with the first
(modest) SFR peak occurred between 5 and 9 Gyr ago. There was star
formation taking place already 13 Gyr ago, but at a very low rate. The
activity in the last 3 Gyr is very low and concentrated in few
episodes.
Figure \ref{cmd_8} shows the synthetic CMD (right panel) generated
from the Bologna solution compared to the data (left panel). The TO
region is well reproduced, as well as the MS, SGB and lower RGB. The
only visible differences are noted at the very bright end of the CMD
($14<V<17$), where a few objects are not matched by any simulation and
are probably foreground stars, and at the very faint end (below
$V=25$), where our model overpredicts star counts.

Despite these similar rates and timings, Cole's and Bologna AMR/CMF
show some slight differences. As found for SFH5, Bologna's metallicity
is systematically higher at early epochs (although still within the
uncertainties). On the other hand, in the range 5-10 Gyr ago Cole's
CMF rises generally faster than Bologna's.

There is also the possibility that the poorly populated upper MS in
SFH8's CMD be contaminated (or dominated) by blue stragglers. It is
well known that such objects populate dwarf spheroidals (see
e.g. \citealt{momany2007}). Given the relatively low densities in
those stellar systems, it is likely that their blue stragglers stem
from primordial binary systems rather than from collisional binaries
as in globular clusters. In this regard, we expect that genuine young
MS stars are likely concentrated on the scale of the star forming
regions while blue stragglers are more widespread, presumably
following the distribution of the bulk of stars in the
SMC. Unfortunately, our field of view is small ($\sim$ 60 pc) compared
to the size of the SMC, so it is virtually impossible to distinguish
any difference in the spatial distribution. Forthcoming wide field
observations with the VST will help elucidate this point.

\subsection{SFH9  \label{sec-sfh9}}

\begin{figure*}
\centering
\includegraphics[width=12cm]{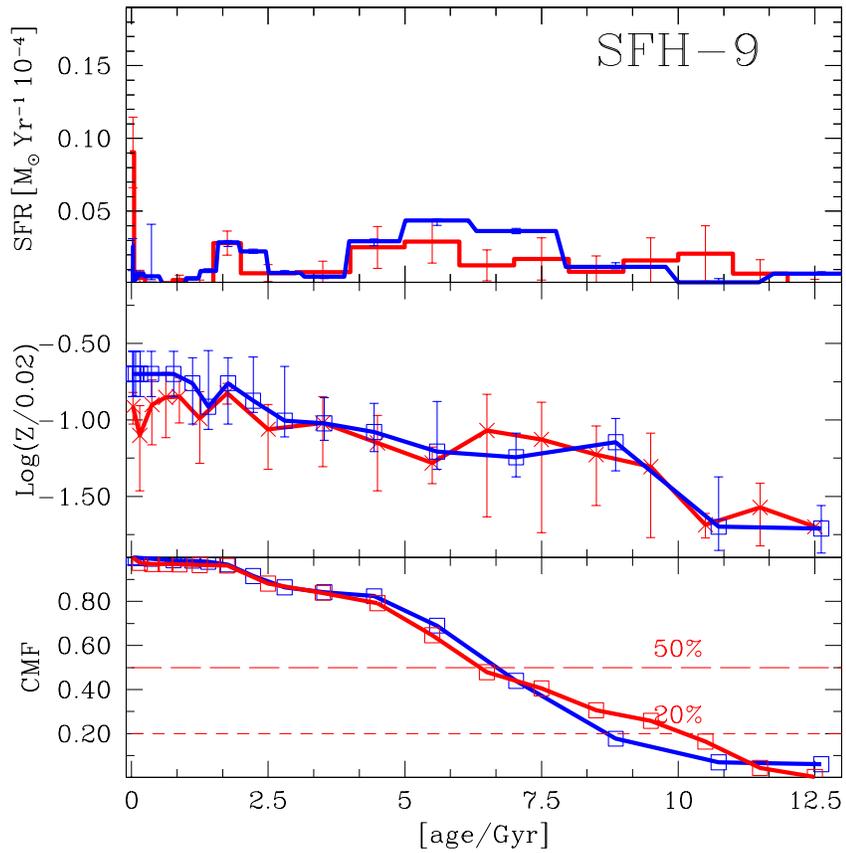}
\caption{SFH, AMR and CMF for SFH9. Colours and lines are as in Figure
  \ref{sfr5}.}
\label{sfr9} 
\end{figure*}

In spite of its large apparent distance from the galactic center and
of its low stellar density, SFH9 contains a significant fraction of
young stars, as clearly indicated by the CMD blue plume. This is
robustly confirmed by our synthetic CMD analyses, which all show a
significant enhancement in the SF rate at very recent epochs (see the
top panel in Fig. \ref{sfr9}).  The only visible differences between
the two solutions are the slightly higher activity of SFH9-C between 5
and 7.5 Gyr ago and the stronger peak of SFH9-B in the last 50
Myr. The corresponding AMRs are in very good agreement, except in the
last 2 Gyr where Cole's metallicity is slightly higher.

In both SFH9-B and SFH9-C the average SFR has been quite low all over
the Hubble time, not much different from that in SFH8, including the
very modest peak around 5-6 Gyr ago and the almost quiescent initial
phases.  What makes this region different from SFH8 is the activity
over the past $\approx$200~Myr.  These stars have not had time to
diffuse throughout the galaxy since their formation, and remain close
to their birthplace in the SMC wing.

\begin{figure*}
\centering
\includegraphics[width=12cm]{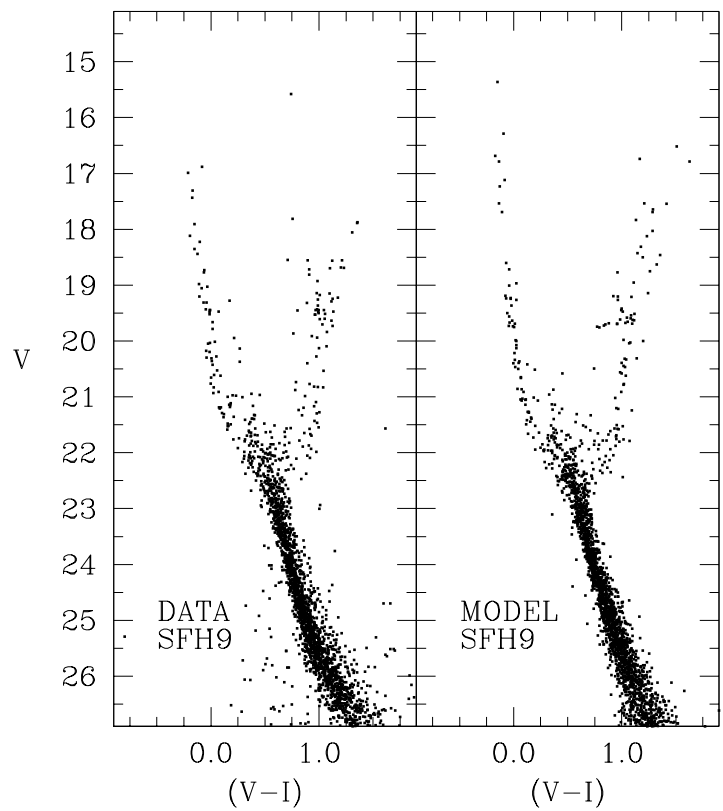}
\caption{Comparison between the observational (left panel) and
  the Bologna synthetic (right panel) CMD for SFH9.}
\label{cmd_9} 
\end{figure*}

Figure \ref{cmd_9} shows the Bologna synthetic CMD (right panel)
compared to the data (left panel). The overall agreement is excellent,
with only minor differences in the RGB, which is sharper in the
synthetic CMDs, and in the He burning region, where our model predicts
a mild HB instead of the observed round RC. It is also worth noting
the apparent broadening of the lower MS, a feature which is not
reproduced by our model. Likely explanations are: 1) the fraction of
binaries in the field SFH9 is larger than the adopted value ( 30\%);
2) the metallicity of the youngest populations is higher than the
expected value; 3) a population of low mass PMS (a stellar phase not
included in these models) stars is present in the field. Since such MS
splitting is not seen in SFH8, whose population is slightly older but
very similar to that of SFH9, we consider the first hypothesis
unlikely.

\subsection{SFH10  \label{sec-sfh10}}

\begin{figure*}
\centering
\includegraphics[width=12cm]{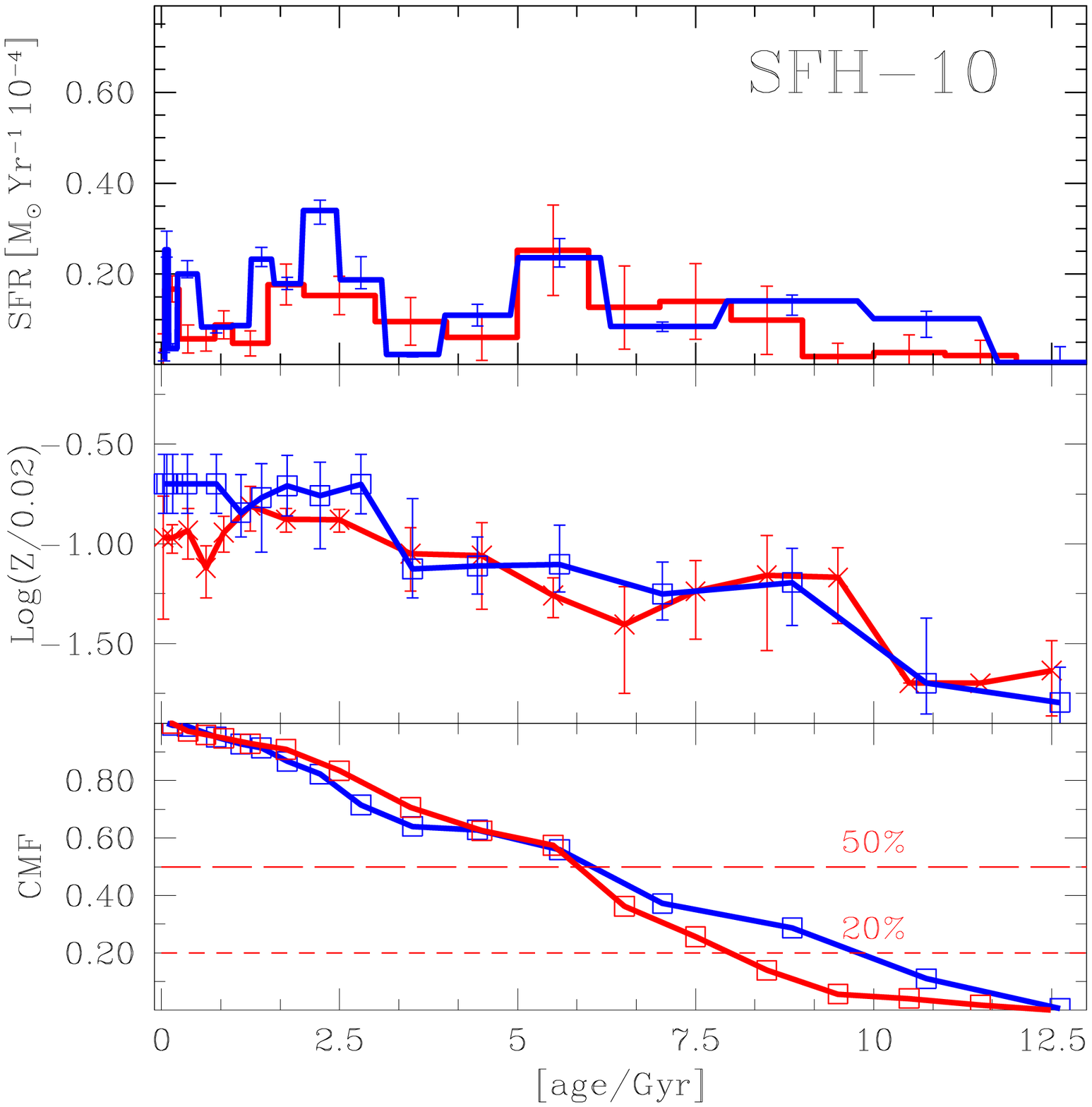}
\caption{SFH, AMR and CMF for SFH10. Colours and lines are as in Figure
  \ref{sfr5}.}
\label{sfr10} 
\end{figure*}

SFH10 lies midway between the bar and the wing, and its SFH is
correspondingly intermediate between SFH5 and SFH9. Figure \ref{sfr10}
shows the resulting SFHs according to the Bologna and Cole
approaches. SFH10-B and SFH10-C agree very well in predicting: 1) a
low activity in the first 5 Gyr; 2) a smooth increase from 8 to 5 Gyr
ago; 3) a hiatus in star formation from 3 to 4 Gyr ago; 4) a SF peak
between 3 and 2 Gyr ago; 5) a fairly smooth decrease since then,
broken by a recent burst of star formation. Both the AMRs and CMFs are
in excellent agreement.  The only difference here is the average rate
of SFH10-C, which is slightly higher than SFH10-B.

The comparison between the SFH9 and SFH10 histories shows intriguing
similarities and differencies.  First of all, the global morphology of
their SFHs is rather similar, although SFH10 has experienced a much
more intense star formation activity for most of the time. However, in
spite of the globally higher rate, the activity of SFH10 in the last
50 Myr is lower than in SFH9. This means that SFH9 has been forming
stars at a slower pace than SFH10 for most of its life as expected for
its larger distance from the SMC center, but has suddenly undergone a
significant SF enhancement that has made it currently much more active
than SFH10. This strongly suggests that the ongoing star formation in
SFH9 is not motivated by the typical gas density in the SMC periphery,
but has been stimulated by some external process. If we add the
evidence that SFH9 lies in the wing region characterized by a string
of HII regions that extend into the Magellanic Bridge, one is driven
to conclude that the interaction with the LMC is the main culprit.

\begin{figure*}
\centering
\includegraphics[width=12cm]{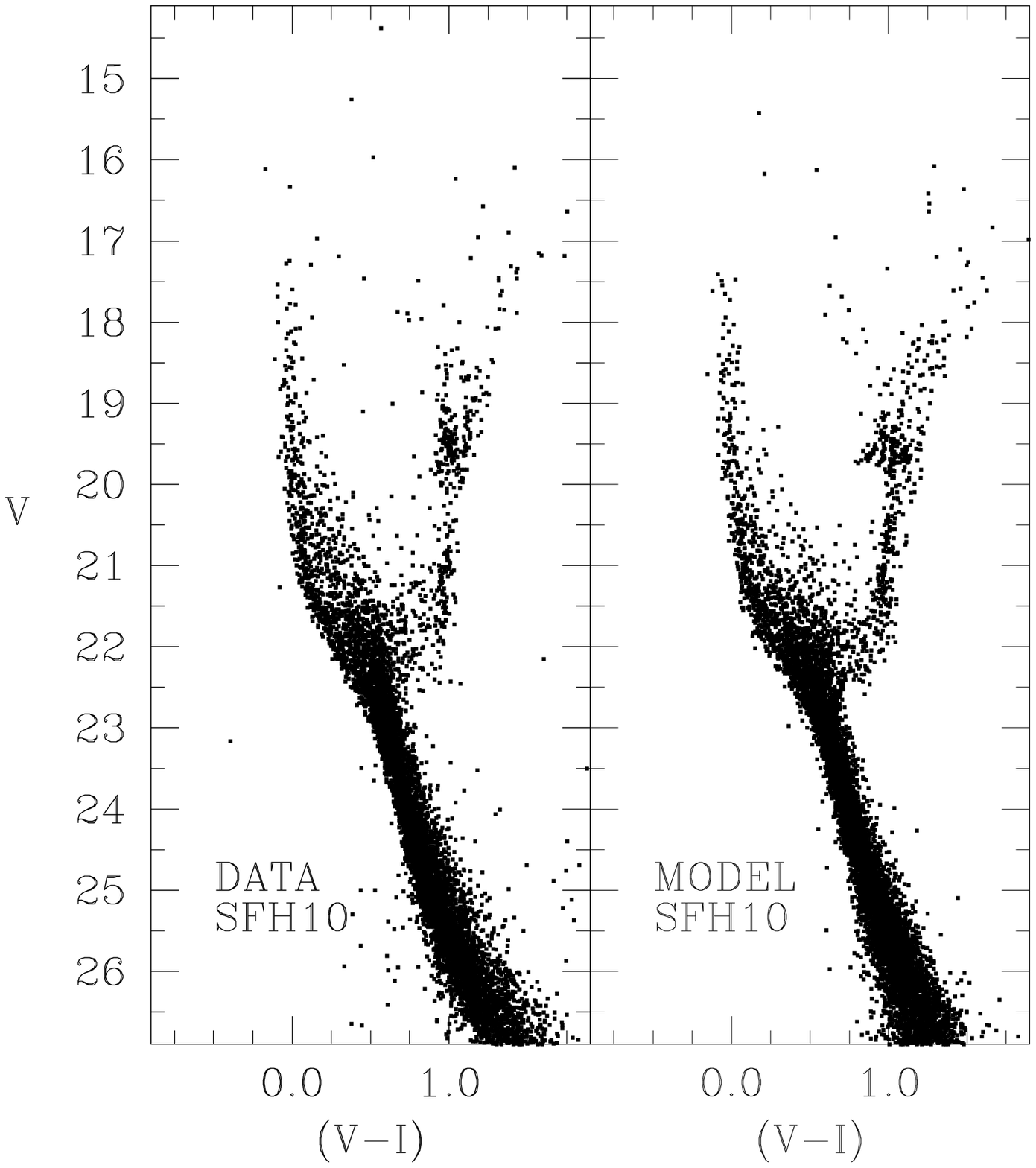}
\caption{Comparison between the observational (left panel) and
  the Bologna synthetic (right panel) CMD for SFH10.}
\label{cmd_10} 
\end{figure*}

Figure \ref{cmd_10} shows the Bologna synthetic CMD (right panel)
compared to the SFH10 data (left panel). Also for this field the model
fits very well the observations. Few minor mismatches remain in the
shape of the RC, which is slightly broader in the model.

\subsection{Age-Metallicity Relation: a global view}

\begin{figure*}
\centering
\includegraphics[width=15cm]{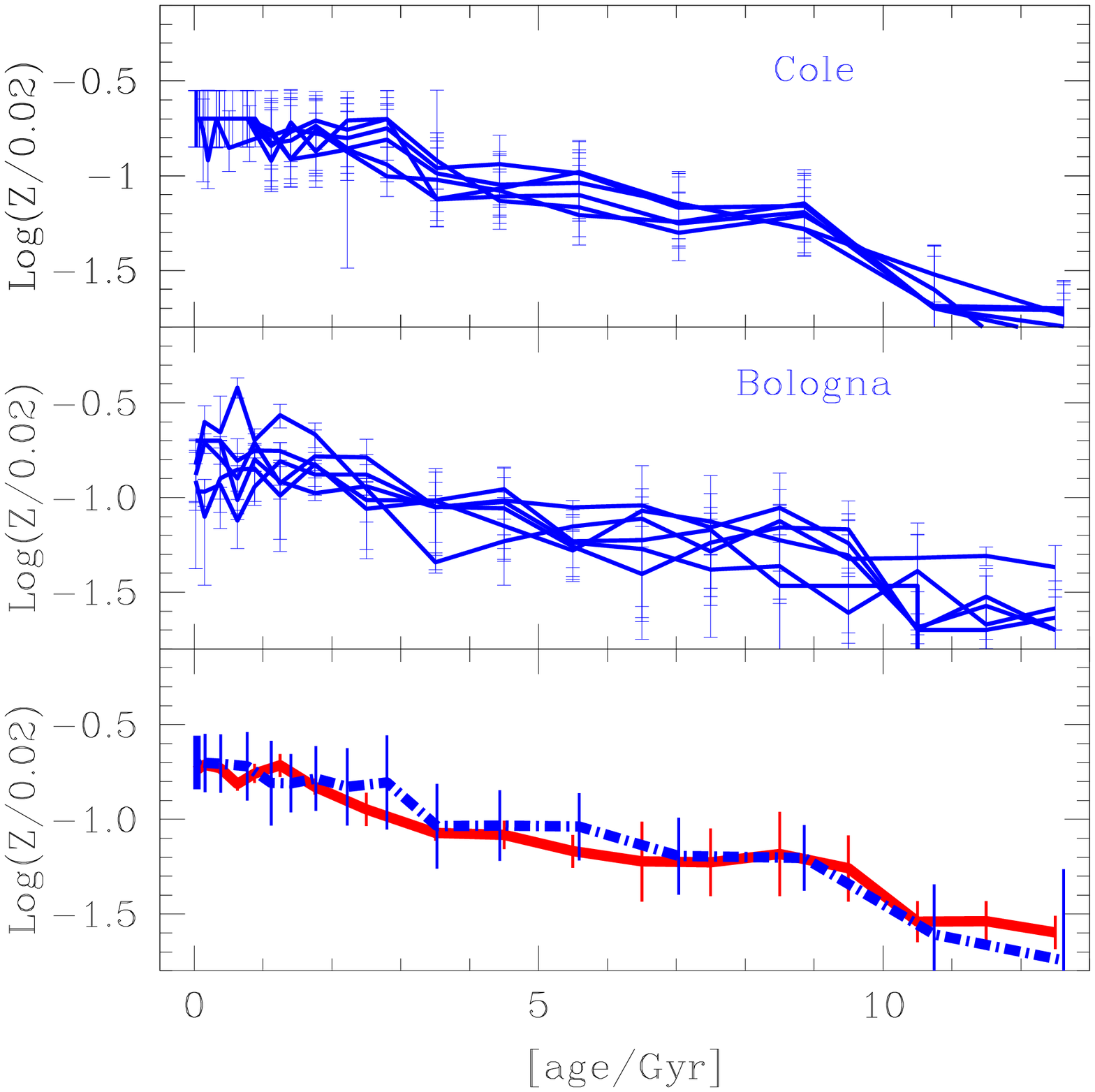}
\caption{Top panel: Cole AMRs. Middle panel: Bologna AMRs. Bottom
  panel: average Bologna AMR (red solid line) vs Cole (dot-dashed blue
  line) AMR.}
\label{amr_bo_co} 
\end{figure*}

In spite of the uncertainties, our AMRs portray a consistent picture
of a metallicity increasing with time.  Figure \ref{amr_bo_co} shows a
composite plot with all our six AMRs (including our solutions for SFH1
and SFH4, see \citealt{cignoni12}). The top and middle panels
summarize Cole and Bologna solutions for each field, while the bottom
panel compares the average Bologna (red line) and Cole (blue line) AMRs.

By comparing the results of the two methods we can conclude that: 1)
within the uncertainties, our solutions are consistent in all fields;
2) the SMC experienced a very slow metallicity evolution until around
2 Gyr ago and a steeper enrichment since then.

There are also few systematic differences that are worth
discussing. At early times the mean metallicity of Bologna solutions
is higher than Cole's, while at intermediate ages it is generally
lower. Moreover, the Bologna solutions show a higher dispersion. All
these variations are mostly due to differences in the metallicity grid
adopted in the two methods. Given the coarser metallicity resolution
of the Bologna set of stellar models, and because of the degeneracy
between reddening and metallicity, a small variation of the adopted
reddening may result in a quite different average metallicity (up to
0.3 dex). This effect is necessarily exacerbated in those fields which
are dominated by an old population. Moreover, the lowest metallicity
of the Bologna set of stellar models (Z=0.0004) is higher than the
corresponding one of Cole's set (Z=0.00015): this naturally explains
the lower metallicity predicted at early times by Cole's method.

%In literature there are two categories of SMC AMRs, those derived
%from field stars and those derived from clusters. Assessing to what
%extent these two represent the same physical process is still largely
%debated.

\section{Comparison with other studies}

\subsection{SFHs}

Our results confirm that most of the central SMC regions have had very
little star formation activity in the first few billion years of
galaxy life, as already found by several authors
\citep{dolphin01,mccumber05,chiosi07,cignoni12, weisz13}. Further
support to this conclusion is provided by the relatively low number of
RR Lyrae stars detected in the SMC compared to the LMC
(\citealt{soszy10}) and by the circumstance that the SMC oldest
cluster, NGC121, is only 11$\pm$0.5 Gyr old \citep{glatt08a}, i.e. much
younger than the oldest globular clusters hosted in the Galaxy and in
the LMC.

\cite{noel07} also found that a shared feature of all their 12 SMC
fields is the absence of a well-populated, blue, extended HB, pointing
out that the amount of field stellar populations as old and metal-poor
as that of the MW Halo globular clusters and dwarf spheroidal galaxies
is small in the SMC. However, in their subsequent synthetic CMD
analysis, \cite{noel09} suggested that a significant SF activity took
place also at the earliest epochs, with a sizable difference between
eastern and western fields.

\citet{harris04}, who were the first to apply the synthetic CMD method
to the derivation of the SMC SFH, suggested that 50\% of SMC stars
formed earlier than 8.4 Gyr ago, and that very few formed in the
period between 3 and 8.4 Gyr ago.  However, contrary to all the
subsequent quoted studies, their ground-based photometry did not reach
the oldest MSTO and this hampered the derivation of the SFH at
relatively early epochs.  This points to the importance of high
angular resolution in minimizing the effect of crowding on deep
color-magnitude diagram analyses. We will be able to further quantify
the star formation level at the earliest epochs over the full extent
of the SMC when the VST Guaranteed Time Observations are completed,
but these must still be tied to diffraction-limited imaging studies
where the stellar density is high.

All our fields experienced their first significant star formation
activity around 8-10 Gyr ago, which reached a peak a few Gyr
later. This behavior is shared also by the fields studied by
\cite{dolphin01}, \cite{mccumber05}, \cite{chiosi07}, \cite{noel07},
and \cite{noel09}.

As displayed by \cite{cignoni12} in their figure 1, our examined
regions lie close to some of those analyzed by other authors and it is
interesting to compare our results with theirs.

Our field SFH8 in the northern outskirts of the galaxy is rather close
to those studied by \citet{dolphin01}, who concluded that stars in the
outskirts of the SMC formed during a broadly peaked episode of star
formation, with the largest (although moderate) rate between 5 and 8
Gyr ago. We find exactly the same result. SFH8 is also not too distant
from the field qj0033 studied by \cite{noel09}, who derived for it a
SFH consistent with ours, with two moderate activity peaks 5 and 8 Gyr
ago. The only significant difference is before 10 Gyr ago, where
\citeauthor{noel09}'s \citeyearpar{noel09} activity is higher than in
our solutions. However, most of the information at these epochs is
conveyed by the oldest TOs, which are much better defined in our data.
\citet{weisz13} also studied a region near our SFH8, reanalyzing WFPC2
fields (their fields 4--7).  They find a similar truncation of major
star formation at ages $\approx$3--5~Gyr, with a median formation age
of $\approx$7~Gyr, identical to our result within the errors.

On the opposite side of the SMC, our region SFH9 is the most external
field studied so far with ACS. The closest field available in the
literature with SFH inferred from the CMD is \cite{noel09} qj0116,
located half way between our SFH9 and SFH10, and the two analyses are
in excellent agreement.  In SFH9 we find a moderate SF activity,
characterized by a very recent burst, a moderate peak around 5--6 Gyr
ago, and very low rates in the first 4 Gyr. In qj0116 \cite{noel09}
also found that the most significant star forming activity was in the
last 1 Gyr, preceded by a more modest rate over most of the Hubble
time, with a secondary peak 5 Gyr ago and a very low initial rate.

Our intermediate field SFH10 lies close to the field studied by
\citet{mccumber05} and the three fields analyzed by \cite{noel09}; in
particular SFH10 is near their qj0111 field.  Our analysis shows that
SFH10 has had a fairly continuous SFH in the last 8-9 Gyr, after the
usual enhancement over the very modest initial activity. This is only
partially consistent with \citet{mccumber05} who found for their
region an increasing rate from 12 to 4 Gyr ago and over the past 1.5
Gyr, with a significantly quieter phase between 4 and 1.7 Gyr
ago. \cite{noel09} found in qj0111 a highly variable SFH, with a first
secondary SF peak 10 Gyr ago followed by a dip 3 Gyr later, a primary
peak about 4 Gyr ago followed by a similar dip 2 Gyr ago and a new
recent peak. Although a similar ``gasping'' behavior is also found in
our solution, the timing of the peaks is different and the earliest
activity is higher than ours. In conclusion, none of the three studies
is in good agreement with the other two.

SFH5 is not coincident with any field with SFH derived from the CMD,
except of course those by \cite{harris04} that cover the whole SMC,
but suffer from poor completeness affecting the SFR measurements for
ages older than a few Gyr.  Of the fields with photometry reaching the
old MSTO, smc0057 of \cite{noel09} is not too distant, and field SMC-1
of \citet{weisz13} is nearly midway between our SFH1 and SFH5. For
smc0057 \cite{noel09} suggest a bouncing SFH not too different from
what we find for SFH5, except that in our field we have a significant
recent burst which is absent in their field, probably because theirs
is slightly more external and away from the star forming
regions. Another interesting difference is that also in this region we
find only a very moderate initial activity, while smc0057 shows a
primary peak 12 Gyr ago. 

\citet{weisz13} find evidence for a burst of star formation
$\sim$9~Gyr ago in their field SMC-1, with a long period of low SFR
punctuated by a minor episode at 5~Gyr and then a dramatic and
sustained increase at $\sim$3~Gyr. In this respect \citet{weisz13}
solution is a more extreme version of our Cole solution, sharing the
same median age of formation with that solution (Fig.\ 3). The reason
for the differing SFR at old age is not obvious, as the WFPC2
completeness levels are not dramatically shallower than the ACS
levels; however it is worth noting that over the time period from
$\sim$7--13~Gyr ago, the error bars on the CMF shown by
\citet{weisz13} are skewed to the low side, suggesting that the
significance of the increase in SFR at 9~Gyr is low.

Finally, we recall that our SFH1 region, described by
\cite{cignoni12}, almost coincides with two of the three deep bar
fields studied by \cite{chiosi07} around the SMC clusters K~29,
NGC~290, and NGC~265, as well as SMC-2 and SMC-3 from
\citet{weisz13}. Our and their solutions are in good qualitative and
quantitative agreement and show an unambiguous rise of the SFR between
7 and 5 Gyr ago and a very moderate earlier activity.  However,
\citet{weisz13} tends to find that the 20\% level in the CMF is
reached earlier and the 50\% level is reached later than in our
solutions.

%\clearpage
\subsection{AMRs}

\subsubsection{Field stars AMR}

\begin{figure*}
\centering
\includegraphics[width=15cm]{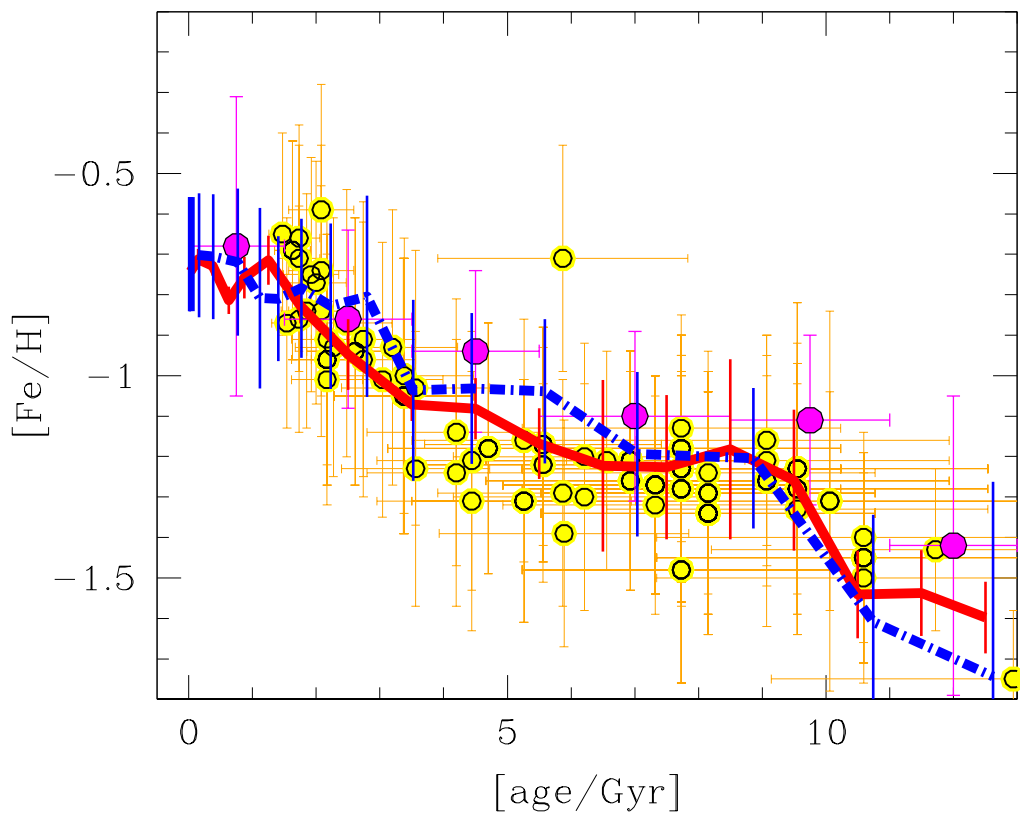}
\caption{Comparison of our predicted average AMRs with literature data
  for SMC field stars. Solid red line: average Bologna AMR. Blue
  dot-dashed line: average Cole AMR. Yellow filled circles: mean ages
  and photometric metallicities of selected fields in the SMC derived
  by \citep{piatti2012a}. Each point represent a SMC sector of about
  $9\times 9$ arcminutes$^2$. Magenta filled circles: Ca II triplet
  AMR derived by \cite{carrera2008}.}
\label{amrs_campo} 
\end{figure*}

Concerning the field AMR, Fig. \ref{amrs_campo} shows Bologna and Cole
solutions\footnote{To convert from Z metallicities to [Fe/H] values we
  adopted Z$_\odot$ = 0.02 and [Fe/H] = log(Z/Z$_\odot$).} along with
the most comprehensive field AMR studies, namely
(\citealt{piatti2012a}, thereinafter P12, yellow filled circles) and
\citeauthor{carrera2008} \citeyearpar[magenta filled triangles,
hereinafter C08]{carrera2008}. These works are independent and
spatially complementary: the former derived a global AMR using
Washington photometry for 160 $9 \times 9$ arcmin$^2$ regions across
the SMC main body, while the latter used Ca II triplet spectroscopy
for 13 regions in the SMC outskirts.
Overall our solutions compare favourably with both P12 and C08 data
\protect\footnote{We point out that the quoted [Fe/H] literature
  values probably adopt different $\log(Fe/H)_{\odot}$ not always
  given in the papers.}, at a level largely consistent with the
uncertainties. The agreement is better when the metallicity is high,
while some differences appear below $[Fe/H]=-1$. In particular, our
AMRs are at the upper edge of P12 distribution between 4 and 6 Gyr
ago, while they are at its lower edge prior to 9 Gyr ago. Conversely,
both the Bologna and Cole AMR are systematically metal poorer than
that of C08 prior to 5 Gyr ago, although always within the error bars.

Taken at face value, these findings might suggest that our HST fields
experienced a chemical enrichment at early times slower than in the
rest of the SMC. The situation is less clear at intermediate ages
where P12 and C08 data enclose our solutions. However, it should be
stressed that the large uncertainties in the ages of both P12 and C08
datasets can contribute to flatten their AMR, e.g. producing more
relatively metal rich old stars.

\subsubsection{Star cluster AMR}

\begin{figure*}
\centering
\includegraphics[width=15cm]{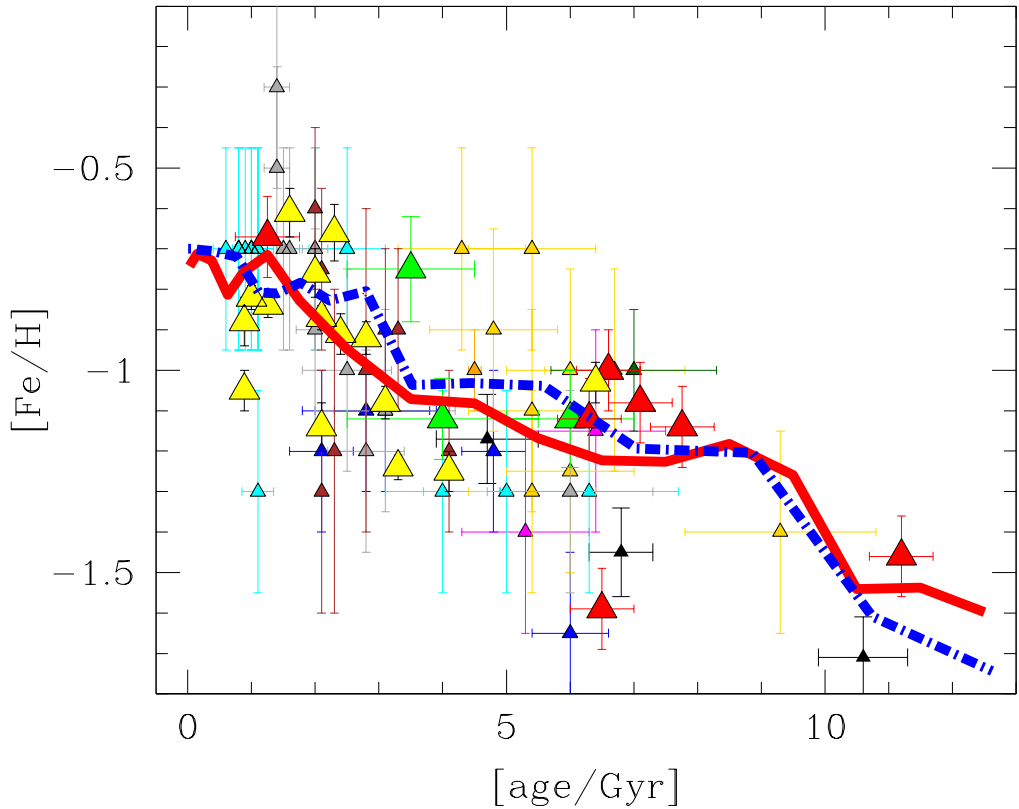}
\caption{Comparison of our predicted average AMRs (same symbols as in
  Fig. \ref{amrs_campo}, theoretical error bars have been omitted for
  clarity) with literature data for SMC clusters. Small and large
  triangles represent photometric and spectroscopic (Ca II triplet)
  determinations, respectively: \cite{dac98} (green filled triangles),
  \cite{glatt08a,glatt08b} (red filled triangles), \cite{parisi09}
  (yellow filled triangles), \cite{mighell98} (black filled
  triangles), \cite{sabbi07} (orange filled triangles),
  \cite{piatti2001} (blue filled triangles), \cite{piatti2007}
  (magenta filled triangles), \cite{piatti2011a} (cyan filled
  triangles), \cite{piatti2011b} (yellow filled triangles),
  \cite{piatti2012b} (dark green filled triangles), \cite{piatti2011c}
  (grey filled triangles), \cite{piatti2005} (brown filled
  triangles).}
\label{amrs_clusters} 
\end{figure*}

Figure \ref{amrs_clusters} shows the comparison with cluster ages and
metallicities collected from the literature. Specifically, small
filled triangles are photometric determinations (\citealt{piatti2001},
\citealt{piatti2005}, \citealt{piatti2007}, \citealt{piatti2011a},
\citealt{piatti2011b}, \citealt{piatti2011c}, \citealt{piatti2012b},
\citealt{mighell98}, \citealt{sabbi07}), while large filled triangles
are spectroscopic determinations (\citealt{dac98}\footnote{Lindsay~1,
  Kron~3 and NGC~121 were not included because ages were updated by
  \cite{glatt08a, glatt08b}.}, \citealt{glatt08a, glatt08b},
\citealt{parisi09}).

Despite the large scatter at any given age and also considering the
large uncertainties, our AMRs are in reasonable agreement with the
cluster AMR. Indeed, it is noteworthy that the
\cite{glatt08a,glatt08b} clusters (filled red triangles), whose ages
have been determined from deep HST CMDs and spectroscopic
metallicities, are those providing the best match to our AMRs (with
the exception of Lindsay 38). Actually, a close inspection suggests
that a large fraction of the cluster symbols are located \emph{below}
our AMRs. As pointed out by \cite{glatt08b}, at any given age the SMC
clusters show a range of metallicities that exceeds the spectroscopic
uncertainties, indicating that the SMC was not well mixed.

\subsubsection{Spectroscopic AMR}

\begin{figure*}[]
\centering
\includegraphics[width=15cm]{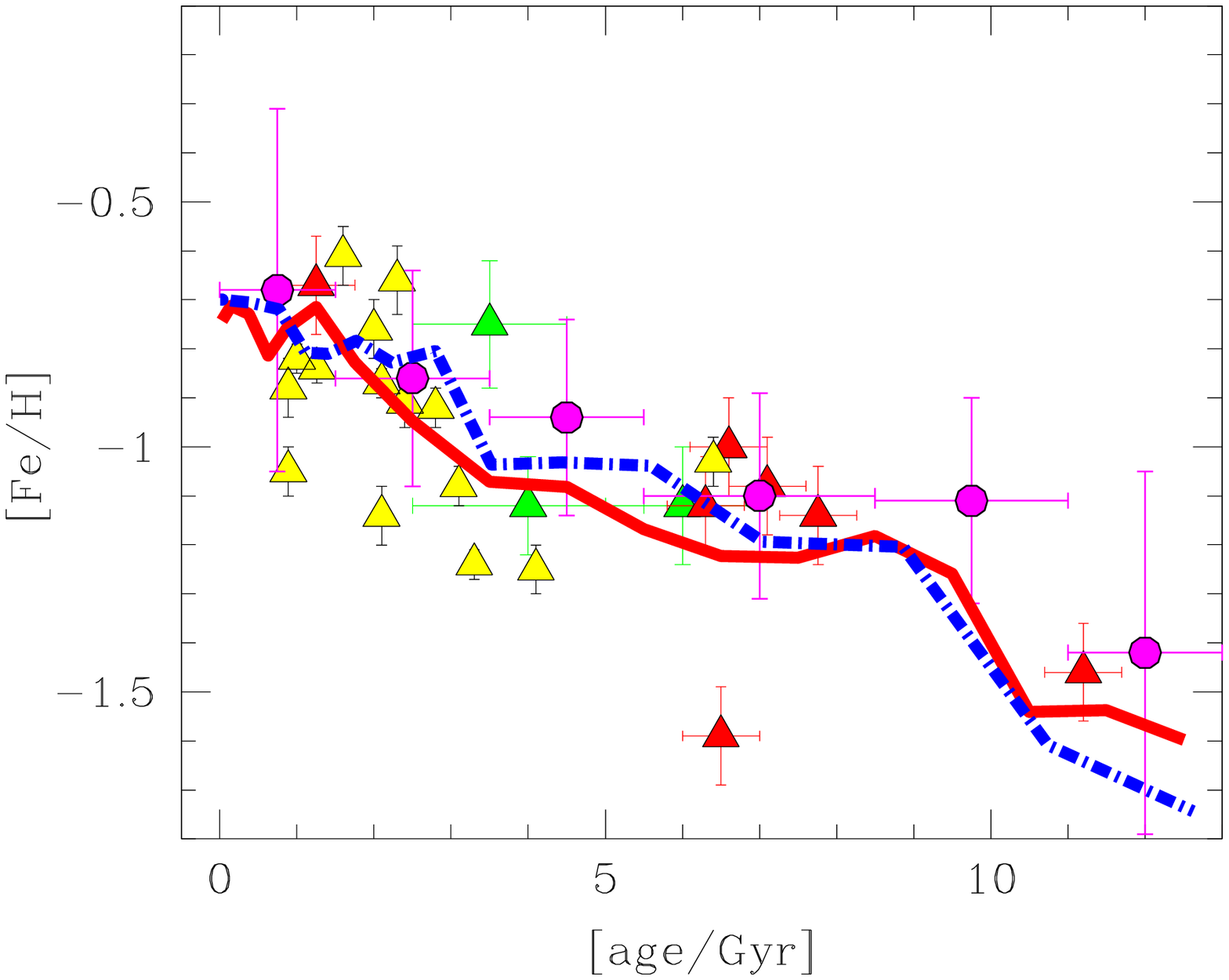}
\caption{Comparison of our predicted average AMRs (same symbols as in
  Fig. \ref{amrs_campo}, theoretical error bars have been omitted for
  clarity) with literature spectroscopic data only (cluster and field
  stars): \cite{dac98} (green filled triangles),
  \cite{glatt08a,glatt08b} (red filled triangles), \cite{parisi09}
  (yellow filled triangles), C08 (magenta filled circles).}
\label{amrs_spectro} 
\end{figure*}

Figure \ref{amrs_spectro} shows the comparison between our AMRs and
all available spectroscopic derivations, regardless if measured in
clusters or in the field. We point out that the general agreement is
improved (in particular between 1 and 9 Gyr ago), with our predictions
within most of observational uncertainties. This suggests that, when
accurate metallicity measurements are taken into account, cluster and
field AMRs may be consistent with each other.

%\clearpage

%A possible reason is that a non negligible fraction of clusters
%was originated from the interstellar matter that fell from the outer
%parts of the SMC (or was captured from disrupted satellite galaxies).

% Taken one at a time, Cole's solution is more satisfactory to
% reproduce the ``knee'' kink at 9 Gyr and the plateau in the range
% 4-8 Gyr, while Bologna enrichment rate is generally too smooth.

\section{Summary and Discussion}

We can summmarize our results as follows:

\begin{enumerate}

\item All six SFHs show that the SMC experienced a global peak of star
  formation between 4-7 Gyr ago.  The onset time of this event is
  consistent across all fields, while the amplitude strongly varies.
  There is some evidence that the duration of the global peak is
  longer in the inner fields, although the degree to which this is a
  continuous process as opposed to a discrete series of shorter events
  cannot be unambiguously assessed with the current data.

  This result is consistent with the sudden appearance of the relative
  excess of clusters found by \cite{piatti2011c} around 7-8 Gyr
  ago. From a theoretical point of view this enhancement poses a
  serious challenge to current dynamical models if we assume that
  cluster and field star formation are primarily
  interaction-triggered. \cite{besla2012} predict that the LMC and the
  SMC are a pair of tidally interacting galaxies that have recently
  been accreted by the MW, while \cite{diaz2011} argue that around 5.5
  Gyr ago the LMC and SMC were isolated (200 kpc of each other). So
  how can we explain the relatively old star formation onset in terms
  of mutual interactions SMC/LMC/MW?

  An intriguing and alternative scenario is the one proposed by
  \cite{tsujimoto2009}: a major merger event took place 7.5 Gyr ago in
  a small group environment that was far from the MW and contained a
  number of small gas-rich dwarfs comparable to the SMC. Although
  attractive, their model also predicts a dip in the SMC AMR (due to
  the large gas infall during the merging), which is in contrast with
  our solutions showing flat or slightly increasing profiles. However,
  a minor merger, which would produce a small dip, may be still a
  viable possibility.
  
  It should be kept in mind that major interactions are not needed for
  small galaxies to experience a sudden increase in SFR following a
  long period of inactivity-- examples include IC~1613
  \citep{skillman03}, DDO~210 \citep{mcconn2006}, Leo~A
  \citep{cole07}, IC~10 \citep{hunter01,cole10}, and NGC~6822
  \citep{cannon12}.  The latter two are of very similar total mass to
  the SMC, and while they are both undergoing a current episode of
  interaction-triggered star formation, they have similar mean ages to
  the SMC and no obvious counterpart for a major interaction.  This
  may be a hint that SMC-mass galaxies are capable of large excursions
  in mean SFR without major mergers or interactions.

\item The sequence of fields SFH1, SFH5, SFH4, SFH10, SFH9 and SFH8
  represents a sequence of age from the youngest (SFH1) to the oldest
  (SFH8). This general trend reflects well the distribution of star
  clusters (see Figure 7 in \citealt{glatt10}).

  All fields share the common characteristics of having formed less
  than 20\% of their mass prior to 10 Gyr ago.  The median age rises
  from $\approx$4 to $\approx$6 Gyr as the projected distance
  increases from 0.05 to 2.2~kpc, largely owing to the decreasing
  amplitude of the intermediate-age SFR enhancement that dominates the
  bar fields. The entire CMF appears to be shifted to older ages in
  the outer fields, with the 20th percentile of stellar mass in place
  by $\approx$9.5~Gyr in the fields at $\sim$2.3~kpc, but not until
  $\sim$5--6~Gyr in the bar.  The bar fields are indistinguishable
  from the central fields off the bar at the same radius over
  virtually their entire lifetime, confirming that the bar is largely
  a ``Population I'' feature \citep[see for
  example][]{wes97,zaritsky2000}.

\item Field SFH9 shows the most peculiar SFH.  It is located
  $\sim$2.3~kpc from the SMC center, at the same radial distance as
  the extremely quiet field SFH8, but on the side of the SMC closest
  to the LMC, in the wing of the SMC. This region is known for its
  disturbed HI morphology and young stellar populations.  Indeed, we
  find that SFH8 and SFH9 have indistinguishable SFH for ages older
  than $\approx$1~Gyr, with a generally low SFR that shows significant
  activity only for ages older than $\sim$2.5~Gyr.  However, the SFH9
  region hosts a population younger than $\sim$200~Myr that is
  completely absent in SFH8.  Such a population is also far weaker in
  the SFH10 field, which is at a similar position angle but midway
  between the wing and the bar.  The shape of the upper main sequence
  in SFH9 is consistent with all of the bright stars having been
  formed in a single event, as the main-sequence is extremely narrow
  and shows little evidence for a continuous distribution of
  main-sequence turnoffs as exemplified by e.g. SFH5.

  These findings support two mechanisms of star formation: a
  spontaneous mode which depends on the density of cold gas available
  to form stars and a triggered mode induced by the LMC/MW
  gravitational wells. The first scenario can explain the
  progressively lower activity moving away from the SMC center where
  the gas supply was plausibly higher. In the second scenario the
  young stellar populations in the wing field SFH9 are likely
  generated from gas pulled out from the SMC during a recent collision
  with the LMC. As a result the recent activity in SFH10 is probably
  driven by a combination of factors, a relatively high gas supply
  (SFH10 is not much further away than SFH5) and the LMC compression,
  while the activity in SFH9, which is located far away from the SMC
  center, may be totally triggered, producing the strong contrast with
  the total lack of activity in the SFH8 field located at similar
  distance but on the anti-LMC side of the galaxy.

\item Our six AMRs are consistent with each other and taken as a whole
  are consistent with a fast initial enrichment prior to 9 Gyr ago,
  very slow metallicity evolution from $\approx$4--9 Gyr ago, and a
  second epoch of major enrichment at more recent times. This is in
  agreement with previous photometric (P12) and spectroscopic (C08)
  studies.

  Our six SFHs do not show evidence for a steep metallicity gradient
  in the the SMC. However, the concentration of younger stars toward
  the inner regions combined with the AMR could be consistent with an
  apparent shallow metallicity gradient. Since the metallicity
  increases with time and the mean age increases with radius, the
  typical star at large radius is expected to be more metal-poor than
  the typical star at small radius. However, our CMDs indicate that at
  a given age there is very little difference in mean metallicity as a
  function of radius, a result largely borne out by spectroscopy of
  age-selected stellar tracers (see \citealt{wes97}, Chapter 11, and
  references therein).

  Previous searches for spatial gradients have produced contrasting
  results. The spatial segregation of young and old stars was noted by
  \citep{gard92} among others. Recently \citep{nidever2011} have
  detected a ~6 Gyr old relatively metal-poor population extending
  out to ~8 degrees in radius, while the younger stars (for example,
  carbon stars, \citealt{morgan95}) are largely contained within 4
  degrees of the center. Among the stars and clusters older than ~1
  Gyr there appears to be considerable scatter in metallicity, which
  tends to obscure trends by increasing the shot noise in typical
  samples.  Among recent spectroscopic studies there are significant
  differences in the reported field star metallicity trend with radius
  from 0$^{\circ}$--6$^{\circ}$, including virtually no trend
  \citep{parisi10}, a steep gradient of $\lesssim$0.2~dex deg$^{-1}$
  (C08), and spatially segregated components at $-$0.6 (inner) and
  $-$1.25 (outer; \citealt{depropris10}).  Photometric studies of
  field stars (e.g., P12) detect metal-rich stars
  concentrated toward the central regions but little to no evidence
  for gradients among the older stars. Although it is generally agreed
  that the younger populations concentrate towards the central
  regions, the degree of concentration, the characteristic ages of the
  inner and outer populations, and the steepness of the AMR over the
  relevant timescales are matters of continued discussion.

\end{enumerate}

\acknowledgments MC and MT have been partially funded with contracts
COFIS ASI-INAF I/016/07/0, ASI-INAF I/009/10/0 and PRIN-MIUR 2010
LY5N2T. EKG acknowledges partial funding from Sonderforschungsbereich
"The Milky Way System" (SFB 881) of the German Research Foundation
(DFG), especially via subproject A2. Partial support for JSG's
analysis of data from HST program GO-10396 was provided by NASA
through a grant from the Space Telescope Science Institute, which is
operated by the Association of Universities for Research in Astronomy,
Inc., under NASA contract NAS 5-26555.

%%\clearpage

%\begin{figure}
% \includegraphics[width=8cm]{fig9.ps}
%\caption{Predicted (red histograms) and observed (black
%  histograms) LFs, computed as in Fig. \ref{LFs}, but using 
%  the SFH1-B solution with expanded metallicity range.}
%\label{LFs_allZ} 
%\end{figure}

%\clearpage

\end{document}